\documentclass[prb,aps,twocolumn,showpacs,showkeys]{revtex4-1}
\usepackage{latexsym}
\usepackage[xdvi]{graphicx}
\usepackage{amssymb,amsmath}
\DeclareMathOperator{\Tr}{Tr}
\newcommand{\lk}{\left(}
\newcommand{\rk}{\right)}
\newcommand{\lab}{\left|}
\newcommand{\rab}{\right|}

\newcommand{\lkv}{\left[}
\newcommand{\rkv}{\right]}

\newcommand{\bsl}[1]{\begin{slide}{#1}}
\newcommand{\esl}{\end{slide}}
\newcommand{\be}{\begin{equation}}
\newcommand{\ee}{\end{equation}}
\newcommand{\ben}{\begin{enumerate}}
\newcommand{\een}{\end{enumerate}}
\newcommand{\bit}{\begin{itemize}}
\newcommand{\eit}{\end{itemize}}
\newcommand{\been}{\begin{displaymath}}
\newcommand{\eeen}{\end{displaymath}}
\newcommand{\ba}{\left[\begin{array}}
\newcommand{\ea}{\end{array}\right]}
\newcommand{\bac}{\begin{array}}
\newcommand{\eac}{\end{array}}
\newcommand{\bc}{\begin{center}}
\newcommand{\ec}{\end{center}}
\newcommand{\bea}{\begin{eqnarray}}
\newcommand{\eea}{\end{eqnarray}}
\newcommand{\bean}{\begin{eqnarray*}}
\newcommand{\eean}{\end{eqnarray*}}
\newcommand{\bqu}{\begin{quote}\begin{it}}
\newcommand{\equ}{\end{it}\end{quote}}

\newcommand{\va}{\mathbf{a}}

\newcommand{\vvr}{\mathbf{r}}
\newcommand{\vq}{\mathbf{q}}

\newcommand{\vz}{\mathbf{z}}

\begin{document}
\title{Elastic anisotropy and Poisson's ratio of solid helium under pressure}
\author{A. Grechnev}
\affiliation{B. Verkin Institute for Low Temperature Physics and
Engineering, National Academy of Sciences, 47 Lenin Ave., 61103
Kharkiv, Ukraine}
\author{S. M. Tretyak}
\affiliation{B. Verkin Institute for Low Temperature Physics and
Engineering, National Academy of Sciences, 47 Lenin Ave., 61103
Kharkiv, Ukraine}
\author{Yu. A. Freiman}
\affiliation{B. Verkin Institute for Low Temperature Physics and
Engineering, National Academy of Sciences, 47 Lenin Ave., 61103
Kharkiv, Ukraine}
\author{Alexander F. Goncharov}
\affiliation{Geophysical Laboratory, Carnegie Institution of Washington, 5251 Broad Branch Road NW, Washington DC 20015, USA}
\affiliation{Center for Energy Matter in Extreme Environments and Key Laboratory of Materials Physics, Institute of Solid State Physics, Chinese Academy of Sciences, 350 Shushanghu Road, Hefei, Anhui 230031, China}
\author{Eugene Gregoryanz}
\affiliation{School of Physics and Centre for Science at Extreme Conditions, University of Edinburgh, Edinburgh EH9 3JZ, UK}
\begin{abstract}
The elastic moduli, elastic anisotropy coefficients, sound velocities and Poisson's ratio
 of hcp solid helium have been calculated using density functional theory
in generalized gradient approximation
(up to $30$ TPa), and pair+triple 
semi-empirical potentials (up to 100 GPa). Zero-point vibrations
have been treated in the Debye approximation assuming $^4$He isotope
(we exclude the quantum-crystal region at very low pressures from consideration).
Both methods give a reasonable agreement
with the available experimental data. Our calculations predict significant elastic
anisotropy of helium ($\triangle P \approx 1.14$,
$\triangle S_1 \approx 1.7$, $\triangle S_2 \approx 0.93$ at low pressures).
Under terapascal pressures helium becomes more elastically isotropic.
At the metallization point there is a sharp feature
in the elastic modulus $C_S$, which is the stiffness with respect to the
isochoric change of the $c/a$ ratio. This is connected with the previously
obtained sharp minimum of the $c/a$ ratio at the metallization point.

Our calculations confirm the previously measured decrease of the Poisson's ratio with increasing
pressure. This is not a quantum effect, as the same sign of the pressure effect was obtained
when we disregarded zero-point vibrations.
At TPa pressures Poisson's ratio reaches the value of $0.31$ at the theoretical
metallization point ($V_{mol}=0.228$ cm$^3$/mol, $p=17.48$ TPa) and $0.29$ at 30 TPa. 
For $p=0$ we predict a Poisson's ratio of
$0.38$ which is in excellent agreement with the low-$p$-low-$T$ experimental data.
\end{abstract}

\keywords{rare-gas solids, solid helium, elastic moduli, Poisson's ratio,
 DFT calculations\\
E-mail: shrike4625@yahoo.com}

\maketitle

\section{Introduction}

Helium is the second element in the periodic table, as well as the second most abundant chemical
element in the universe. It is a major constituent of both stars and giant planets, and it
is involved in several different nuclear fusion reactions. 
Its high-pressure physical properties are therefore rather important for many
different branches of natural science. The low-temperature behaviour
of helium is well-known but rather peculiar: it stays liquid up to the absolute
zero while becoming superfluid, quantum-freezes under pressure, and demonstrates quantum-crystal
behavior in the solid phase. These quantum 
effects are a consequence of a relatively small mass of the He atom
($\approx 7296$ electron masses for $^4$He) and weak interatomic interaction.
Solid helium is a close-packed atomic crystal
with nearly spherically symmetric He atoms. $^4$He has hexagonal close-packed (hcp)
structure everywhere except for small body-centered cubic (bcc) and face-centered cubic (fcc)
regions near the melting line on the $(p,T)$ phase diagram \cite{loubeyre93prl71:2272}.
The $c/a$ ratio of the hcp structure is close to the ideal value $\sqrt{8/3} \approx 1.633$
\cite{freiman09prb80:094112,grechnev10fnt36:423}.

While helium has been an object of intensive experimental study for more than a century,
the high-pressure experimental breakthrough happened in the last few decades due to the invention
of diamond anvil cells. The elastic moduli of hcp helium up to 32 GPa, as well as
sound velocities, Poisson's ratio (PR) and elastic anisotropy parameters, have been measured 
experimentally by Zha et al. \cite{zha04prb70:174107}.
The results were somewhat
unexpected. First, a significant anisotropy of elastic properties was found.
The problem of the elastic anisotropy of helium is important for the high-pressure
experimental techniques, as helium is frequently used as a quasi-hydrostatic
medium \cite{takemura01jap89:662}. Second,
the Poisson's ratio was found to decrease with increasing pressure, which is a rather unusual
behavior, as for most solids PR increases with pressure, approaching $1/2$ at megabar pressures.
Similar decrease of PR with pressure has been observed in solid hydrogen \cite{zha93prb48:9246}.
Such anomalous PR behavior of He was often thought to be
a quantum zero-point vibration (ZPV) effect, however, as we
show in the present paper and in Ref. \onlinecite{freiman15fnt41:571},
this is not the case.
A theoretical calculation of elastic moduli by Nabi et al. \cite{nabi05prb72:172102} using
density functional theory (DFT) in local Airy gas (LAG),
local density approximation (LDA) and  generalized gradient approximation (GGA)
soon followed the experiment.  The calculated elastic moduli 
were in a reasonably good agreement with experiment. Unfortunately, the authors of
Ref. \onlinecite{nabi05prb72:172102} did not calculate
Poisson's ratio, and did not study the elastic anisotropy in any detail either. 

The goal of the present paper is to clarify the two issues introduced above.
We calculate the elastic moduli of hcp He as a function of pressure
using two complementary methods: DFT-GGA and semi-empirical (SE) potentials, specifically
focusing on the Poisson's ratio and elastic anisotropy parameters. We extend our GGA 
calculations into the metallic phase (up to $p = 30$ TPa)
in order to check the effect of terapascal pressures and the metallization transition 
on the elastic properties of helium. We also
study the effect of zero-point vibrations in the Debye approximation
on the physical quantities in question (for $^4$He)
in order to determine whether quantum effects play any significant role at high
($p \gtrsim 10$ GPa) pressures.

The paper is organized as follows. In section \ref{s:method} we outline the two approached
for calculating total energy (DFT-GGA and SE), and then give a brief introduction to the elasticity
of hexagonal crystals, the algorithm of calculating elastic moduli of an hcp crystal
\cite{steinle99prb60:791}, and define physical quantities used in the present paper.
In section \ref{s:results} we present results of our calculations.
\section{\label{s:method}Method}
\subsection{Total energy calculations}
Helium consists of electrons and nuclei, and due to the relatively small mass of the latter,
their quantum zero-point motion cannot be ignored in general. Many different numerical methods
have been applied to solid helium \cite{mcmahon12rmp84:1607,cazorla15prb91:024103,monserrat14prl112:055504}. 
The electronic subsystem can either be described from first principles, like in density
functional theory (DFT)-based methods, or replaced by empirical pairwise or $n$-body interactions
between nuclei. The quantum and thermal motion of the nuclei can be analyzed either in
harmonic approximation, or using anharmonic approaches, such as diffusion Monte Carlo (DMC)
and other Monte Carlo methods \cite{mcmahon12rmp84:1607}.

For the elasticity calculations pair potentials are not sufficient. We have opted
to use DFT in the generalized gradient approximation (GGA) as our main method. We have also
used pair and 3-body empirical potentials for the $p \lesssim 100$ GPa range. For the
quantum zero-point vibrations we have used a simple harmonic Debye approximation (see below).
While such approach is rather crude, and the use of harmonic approximations for He
atoms has been recently criticized in Ref. \onlinecite{cazorla15prb91:024103},
we chose it as it is computationally cheap and consistent with a full DFT treatment
of the electronic subsystem.

For the density functional theory calculations we have used the all-electron
full-potential linear muffin tin orbital (FP-LMTO) code RSPt \cite{wills:rspt-book}
with the GGA functional of 
Perdew, Burke, Ernzerhof (PBE) \cite{perdew96prl77:3865}. The basis set included
1s, 2p and 3d electrons of helium, with two LMTO basis functions with kinetic energies
$-0.1$ Ha and $+0.1$ Ha respectively per each atomic orbital. 847 $k$-points in the
Brillouin zone have been used. 

For the semi-empirical calculations we have used the pair and triple SE potentials 
described in Ref. \onlinecite{freiman07fnt33:719}. They include Aziz pair potential in the Silvera-Goldman
form and the three-body potential in the Slater-Kirkwood form. These are exactly the potentials used in our previous works
\cite{freiman15fnt41:571,freiman08prb78:014301,freiman09prb80:094112,grechnev10fnt36:423,
freiman13prb88:214501} on helium.
The cutoff radii $R_2=50.2 a$ and $R_3=10.2 a$ were used for pair and triple forces respectively,
with $a$ being the lattice constant.

Both DFT and SE calculations were performed for zero temperature,
and the zero-point vibrations were neglected at first. 
The effect of ZPV was later accounted for in the Debye approximation. 
In contrast to Ref. \onlinecite{nabi05prb72:172102}, we found
the pressure-dependent equilibrium $c/a$ ratios as described in Refs.
\onlinecite{freiman09prb80:094112,grechnev10fnt36:423} and used them for all our calculations.
All our results are well converged with respect to the number of $k$-points (GGA)
and cutoff radii (SE) respectively. 

DFT-GGA and SE can be seen as complementary methods. Semi-empirical potentials usually
work very well for low pressures, but fail for higher pressures where 4-body 
and higher order $n$-body forces become important. For helium the threshold
pressure is of the order of 100 GPa \cite{freiman09prb80:094112,grechnev10fnt36:423}.
GGA, on the other hand, can be inaccurate at low pressures due to poor description
of the van der Walls (vdW) forces. It has been shown \cite{cazorla15prb91:024103} that the
effect of the vdW forces is rather small in the gigapascal pressure range.

\subsection{Elasticity under initial pressure}

This subsection and the remaining part of section \ref{s:method} deal with the elasticity
theory for a hexagonal crystal under pressure and the method of calculating elastic moduli numerically.
It includes the main results of
Refs. \onlinecite{barron65pps85:223,steinle99prb60:791,watt80jap51:1525}, and other works.
This material has never before been gathered in one place, therefore we decided to give a brief
introduction to the method as a whole, presenting all relevant formulas and stressing
some important points.

If a strain is applied to an elastic medium, each point $\vvr$ of the medium 
is shifted to a new position $\vvr^{\prime} (\vvr)$, and we can define tensor $u_{ij}$ as
\be
\label{def-strain}
u_{ij} \equiv  \frac{\partial x_i^{\prime}}{\partial x_j} =
e_{ij} + \omega_{ij},
\ee
where $e_{ij} = (u_{ij}+u_{ji})/2$ and $\omega_{ij} = (u_{ij}-u_{ji})/2$ are
the symmetric strain tensor and the antisymmetric
rotation tensor respectively, and the tensor indices $i,j=1,2,3$
number three Carthesian (not crystal) coordinates.
Summation over repeated tensor indices is assumed.
In the present paper we do not consider rotations,
so we always assume that $u_{ij}=u_{ji}=e_{ij}$ and $\omega_{ij}=0$.
Note that we are not using the Lagrangian strain tensor 
$\eta_{ij} \equiv (u_{ij} + u_{ji} + u_{ki} u_{kj})/2$ in the present paper.
The difference between $e_{ij}$ and $\eta_{ij}$ is important 
under external pressure.

The volume of the strained medium is
\be
\label{def-v}
V^{\prime} = V \det (\delta_{ij}+u_{ij}) = V + \triangle V + O(u^3),
\ee
where
\be
\label{def-delta-v}
\triangle V= V \lk u_{ii} + \frac 12 (u_{ii})^2 - \frac 12 u_{ij} u_{ji} \rk
\ee
is the change of volume up to the second order in $u_{ij}$, and we have used
the identity $\det \hat A = \exp \Tr \log \hat A$ to expand $V^{\prime}$ in
powers of $u_{ij}$.

The elasticity theory for a medium under external stress $\sigma^{(0)}_{ij}$
is rather non-trivial \cite{barron65pps85:223,huang50prs203:178} and there is
no straightforward generalization of the zero-pressure stiffness tensor $C_{ijkl}$.
However, theory simplifies for the case of the  isotropic external pressure
$\sigma^{(0)}_{ij} = - p \delta_{ij}$. The 
stress-strain relation up to the first order in $u_{ij}$ is \cite{barron65pps85:223}
\be
\label{hooke1}
\sigma_{ij} = C_{ijkl}(p) u_{kl} - p \delta_{ij},
\ee
where the rank-four pressure-dependent stiffness tensor
$C_{ijkl}(p)$ (tensor of elastic moduli, 
called $\mathring{c}_{\alpha\beta\sigma\tau}$ in Ref. \onlinecite{barron65pps85:223})
has the same symmetry as the zero-pressure $C_{ijkl}$, namely
\be
C_{ijkl} = C_{jikl} = C_{ijlk} = C_{klij}.
\ee
The elastic energy density up to the second order in $u_{ij}$ is \cite{barron65pps85:223}
\be
\label{hooke2}
\epsilon \equiv \frac EV = \frac 12 C_{ijkl}(p) u_{ij} u_{kl} - p \frac{\triangle V}{V},
\ee
where $\triangle V$ is defined in Eq. (\ref{def-delta-v}).  The terms of the order
$u^2$ in $\triangle V$ are important, as they are of the same order as the first term
in Eq. (\ref{hooke2}). Note that the energy density $\epsilon$ is defined with respect
to the undeformed volume $V$, not $V^{\prime}$. Total energy $E$ gives
adiabatic stiffness tensor, while for the isothermic one the Helmholtz free energy
$F=E-TS$ should be used instead. In this paper
we limit ourselves to the case $T=0$, so there is no distinction between the two.
The equation of motion of the elastic medium
up to the first order in $u_{ij}$  is  \cite{barron65pps85:223}
\be
\rho \frac {\partial^2 u_i}{\partial t^2} = 
C_{ijkl}(p) \frac {\partial^2 u_k}{\partial x_j \partial x_l},
\ee
where $u_i \equiv x^{\prime}_i - x_i$.

When calculating elastic moduli numerically from Eq. (\ref{hooke2}) one must be careful with the
$-p \triangle V/V$ term, as it includes $u^2$ terms. One way of addressing the
problem is to calculate bulk and shear moduli 
separately \cite{mehl93prb47:2493,steinle99prb60:791}. Voigt bulk modulus is 
defined as the bulk modulus under the uniform \emph{strain} (i.e. the lattice
geometry is not allowed to change as the volume changes): 
\be
K_V = - \left. V \frac {d^2 E}{d {V^{\prime}}^2} \rab_{V^{\prime}=V, \: \text{fixed geometry}}
\ee
It can be shown that $K_V(p) = C_{iikk}(p)/9$ and it does not depend on $p$
explicitly.
Reuss bulk modulus is defined as the bulk modulus under
the uniform \emph{stress} (i.e. the lattice geometry is changed as the pressure changes): 
\be
K_R = - \left. V \frac {d^2 E}{d {V^{\prime}}^2} \rab_{V^{\prime}=V, \: \text{relaxed geom.}} =
V \lk \left. \frac{d V^{\prime}}{d p^{\prime}}  \rab_{p^{\prime}=p} \rk^{-1}.
\ee
$K_R$ is equal to $1/S_{iikk}(p)$, where $S_{ijkl}(p)$ is the compliance tensor, the inverse of
$C_{ijkl}(p)$, and again it does not depend on $p$ explicitly. 
The shear moduli are calculated from Eq. (\ref{hooke2}) using strains $u_{ij}$
which are isochoric (volume-conserving) exactly, or in the second order of $u_{ij}$ at least,
so that $\triangle V=0$ in Eq. (\ref{hooke2}) and the elastic energy is proportional 
to the shear modulus in question times $u^2$,
without any additional terms proportional to $p u^2$.

Hexagonal close-packed (hcp) crystal lattice has lattice vectors 
\be
\va_1 = a \lkv \begin{array}{c}
1 \\ 0 \\ 0
 \end{array}\rkv, 
\va_2 = a \lkv \begin{array}{c}
-1/2 \\  \sqrt{3}/2 \\ 0
 \end{array}\rkv, 
\va_3 = c \lkv \begin{array}{c}
0 \\ 0 \\ 1
 \end{array}\rkv,
\ee
where $a$ and $c$ are two pressure-dependent lattice constants. 
Atom positions in the undeformed hcp lattice are
\be
\label{pos-cryst}
(0,0,0) , \; \lk \frac 23, \frac 13, \frac 12 \rk
\ee
in crystal coordinates.
When a uniform strain $u_{ij}$ is applied to the medium, any point $x_i$
changes to $x^{\prime}_i = (\delta_{ij}+u_{ij}) x_j$,
i.e. atoms form a deformed crystal lattice with new lattice vectors
\be
\label{strain-a}
a^{\prime}_i = (\delta_{ij}+u_{ij}) a_j.
\ee
 Special care must be taken when applying strains to a lattice
with more than one atom per unit cell. The expression (\ref{strain-a}) determines only the change
of three lattice vectors under strain, but tells nothing about positions of atoms within the
unit cell. Such positions are not determined by macroscopic elasticity theory and
must be allowed to relax in total energy calculations.
If they change linearly in $u_{ij}$ under strain, this affects
the calculated elastic constants. In other words, to obtain correct result
we must minimize the total energy with respect to all atomic positions
for each applied finite strain. Hcp lattice has
two atoms per unit cell, and their positions are fixed by symmetry for the
undeformed lattice, however for certain strains (like the orthorhombic strain, see below)
they indeed change linearly in $u_{ij}$.

In Voigt notation
\be
(11, 22, 33, 23, 31, 12) \to (1,2,3,4,5,6)
\ee
the stiffness tensor of a hexagonal crystal is specified by
five independent elastic constants $C_{11}$, $C_{12}$,$C_{13}$, $C_{33}$ and $C_{44}$:
\be
C_{ij} = \lkv \begin{array}{cccccc}
C_{11} & C_{12} & C_{13} & 0      & 0      & 0 \\
C_{12} & C_{11} & C_{13} & 0      & 0      & 0 \\
C_{13} & C_{13} & C_{33} & 0      & 0      & 0 \\
0      & 0      & 0      & C_{44} & 0      & 0 \\
0      & 0      & 0      & 0      & C_{44} & 0 \\
0      & 0      & 0      & 0      & 0      & \frac 12 (C_{11} - C_{12}) 
\end{array} \rkv.
\ee
\subsection{Calculation of elastic constants}
If we have any method of calculating the energy of the crystal for given 
volume $V$, $c/a$ ratio and strain $u_{ij}$, like DFT or SE,
we can calculate the five elastic constants numerically by applying several
independent strains in Eq. (\ref{hooke2}). An efficient way to do this for the hcp
lattice has been proposed by Steinle-Neumann \textit{et al.} \cite{steinle99prb60:791}.
As explained above, bulk and shear moduli are calculated separately.

First we calculate the total energy $E(V)$ as a function of volume 
(in practice the molar volume $V_{mol}=V N_A/N$ 
is used, where $N$ is the number of atoms, and $N_A$ is Avogadro's number).
For each volume the energy minimum with respect to the ratio $c/a$ is found,
and $E(V)$ is defined as the energy of this minimum.
The ideal $c/a$ ratio (corresponding to the close-packing of hard spheres)
is $\sqrt{8/3} \approx 1.633$, and for helium the $c/a$ ratio is rather close to the ideal
value \cite{freiman09prb80:094112}. The equation of state (EOS) $p(V)$ and the Reuss
bulk modulus $K_R$ are found as
\be
p= - \frac {dE}{dV}, \quad K_R = -V \frac {dp}{dV} = \frac {d^2E}{dV^2}.
\ee
In practice, the energy $E(V)$ is approximated (via least square fit) by the Rose-Vinet
equation of state \cite{vinet87prb35:1945}, which allows for an accurate numerical differentiation.
For GGA, we were unable to find a single Rose-Vinet fit which would be accurate in both GPa and TPa 
pressure ranges, therefore we had to use two different parametrizations.

By using a diagonal stress $\sigma_{ij}=-(p+\triangle p) \delta_{ij}$,
which is an infinitesimal change of pressure,
we can show that the Reuss bulk modulus is
\be
\label{KR-eq}
K_R = \frac {Q}{C_S},
\ee
where
\be
Q \equiv C_{33} (C_{11} + C_{12}) - 2 C_{13}^2,
\ee
\be
\label{CS-eq}
C_S \equiv C_{11} + C_{12} + 2 C_{33} - 4 C_{13}.
\ee
The corresponding strain (found from Eq. (\ref{hooke1})) is
\be
\label{kr-strain}
u_{ij}= - \frac {\triangle p}{Q}
\lkv \begin{array}{ccc}
C_{33} - C_{13} & 0 & 0 \\
0 & C_{33} - C_{13} &  0 \\
0 & 0 & C_{11} + C_{12} - 2 C_{13}
\end{array} \rkv.
\ee
Eq. (\ref{KR-eq}) is the first equation we use for determining five elastic constants.
From Eq. (\ref{kr-strain}) we can find the logarithmic derivative of $c/a$ 
with respect to volume
\be
\label{R-eq}
R \equiv -\frac {d \log (c/a)}{d \log V} = \frac 1{C_S} (C_{33}+C_{13} - C_{11} - C_{12}).
\ee
This is the second equation we need. An analytic parametrization of $c/a(V_{mol})$ is
used for numerical differentiation as usual.

From now on we are going to use only exactly isochoric (volume-conserving) strains
as explained above.
In order to calculate $C_S$, Ref. \onlinecite{steinle99prb60:791} used an isochoric
$c/a$-changing strain, however it is not necessary, as we can use an equivalent formula
\be
\label{CS-eq2}
C_S = \frac {9}{2V} \lk \frac ca\rk^2 \frac {\partial ^2 E(V,c/a)}{\partial \lk c/a\rk^2}
\ee
taken at the equilibrium $c/a$. We use the energies $E(V,c/a)$ calculated
previously when we looked for the equilibrium $c/a$ for each volume and approximate it by
a fourth order polynomial of $c/a$.
Equation (\ref{CS-eq}) with  $C_S$ from Eq. (\ref{CS-eq2})
is the third equation for the five elastic moduli. We now have three equations for three variables
$C_{11}+C_{12}$ (in this combination only), $C_{13}$ and $C_{33}$. They can be solved
to obtain
\be
\label{modulus1}
C_{11}+C_{12}= 2 K_R + \frac {C_S}9 (2R-1)^2,
\ee
\be
\label{modulus2}
C_{13} = K_R + \frac {C_S}9 (2R-1) (R+1),
\ee
\be
\label{modulus3}
C_{33} = K_R + \frac {2 C_S}9 (R+1)^2.
\ee

We need two more isochoric strains to find $C_{44}$ and $C_{66} \equiv (C_{11}-C_{12})/2$.
$C_{44}$ is found from the monoclinic strain
\be
u_{ij}= \lkv \begin{array}{ccc}
0 & 0 & t \\
0 & \frac {t^2}{1-t^2} &  0 \\
t & 0 & 0
\end{array} \rkv,
\ee
with $\epsilon = 2 C_{44} t^2 + O(t^4)$.  Alternatively, $C_{44}$ for the hcp
lattice can be obtained from the calculated $E_{2g}$ Raman frequency
using the formula \cite{olijnyk00jpcm12:10423}
\be
C_{44} = \frac{m}{4 \sqrt{3}} \frac{c}{a^2} \nu^2_{E_{2g}},
\ee
as we did previously in Ref. \onlinecite{freiman13prb88:214501}. The resulting values for 
$C_{44}$, obtained using these two approaches agree within 1\%. For $C_{66}$ we
use the orthorhombic strain
\be
u_{ij}= \lkv \begin{array}{ccc}
t & 0 & 0 \\
0 & -t &  0 \\
0 & 0 & \frac {t^2}{1-t^2}
\end{array} \rkv,
\ee
which gives $\epsilon =2 C_{66} t^2 + O(t^4) = (C_{11}-C_{12})t^2 + O(t^4)$. Note 
that for this strain the atomic positions change under strain. The position of atom 2, 
which is $(2/3,1/3,1/2)$ for the undeformed lattice, becomes $((1+\lambda)/2,\lambda, 1/2)$,
with parameter $\lambda$ depending linearly on $t$ as $\lambda=1/3 - g t + O(t^2)$.
It means that we have to relax the parameter $\lambda$ for each finite strain $t$ in order to calculate
the correct elastic energy.  In our calculations of $C_{44}$ and $C_{66}$ we have used 
five $t$-points $t=0,\pm 0.005, \pm 0.010$ (for very high pressures
smaller values of $t$ were used) and interpolated the energy $E(t)$ with a fourth
order polynomial in order to find the coefficient before $t^2$. The results
obtained this way are virtually independent on the particular values of $t$
used. Such approach is vital for $C_{66}$, while
for $C_{44}$ the $E(t)$ dependence is almost exactly quadratic in $t$ in a wide range of $t$.
With $C_{44}$ and $C_{66}$ calculated we can finally determine all
five elastic constants of the hcp lattice. The quantities $C_{44}$,  $C_{66}$ and
$C_S/6$ are three different shear moduli for three independent pure shear
(isochoric) deformations. They are all equal for an isotropic solid.

\subsection{Anisotropy parameters}
The three acoustic sound velocities (one compressional and two shear ones)
of the hexagonal lattice are \cite{musgrave:book,zha93prb48:9246}
\be
\label{sc-vel1}
\rho v_P^2 = \frac {A+B}2,
\ee
\be
\label{sc-vel2}
\rho v_{S1}^2 = \frac{C_{11}-C_{12}}2 \sin^2 \theta + C_{44} \cos^2 \theta,
\ee
\be
\label{sc-vel3}
\rho v_{S2}^2 = \frac {A-B}2,
\ee
where
\be
A \equiv C_{11} \sin^2 \theta + C_{33} \cos^2 \theta + C_{44},
\ee
\be
B^2 \equiv \lkv (C_{11}-C_{44}) \sin^2 \theta + (C_{44}-C_{33}) \cos^2 \theta\rkv^2
\ee
\[
+(C_{13}+C_{44})^2 \sin^2 (2 \theta),
\]
and $\theta$ is the angle between wave vector $\vq$ and $\vz$.

The elastic anisotropy of a hexagonal crystal can be described by the
anisotropy parameters of these three acoustic waves \cite{steinle99prb60:791}
\be
\triangle P = \frac{C_{33}}{C_{11}},
\ee
\be
\triangle S_1 = \frac{C_{11}+C_{33}-2 C_{13}}{4 C_{44}} = \frac{C_S + 2 C_{66}}{8 C_{44}},
\ee
\be
\triangle S_2 = \frac{2 C_{44}}{C_{11}-C_{12}} = \frac{C_{44}}{C_{66}}.
\ee
For an isotropic medium this quantities are equal to one. 
Note that for a cubic crystal $\triangle P$  is equal to unity,
and  $\triangle S_1$ is the single
anisotropy parameter $(C_{11}-C_{12})/(2C_{44})$ of the cubic crystal.
$\triangle S_2$ is ambiguous, as there is no condition $C_{66}=(C_{11}-C_{12})/2$
for the cubic symmetry, but instead $C_{44}=C_{66}$.
\subsection{Aggregate properties: Voigt and Reuss approaches}
Aggregate description replaces an actual crystal with an effective
isotropic elastic medium, described by an average bulk modulus $K$ and
an average shear modulus $G$, with
\be
C_{ijkl} = \lk K-\frac 23 G \rk \delta_{ij} \delta_{kl} +
G \lk \delta_{ik} \delta_{jl} + \delta_{il} \delta_{jk}\rk.
\ee
It is a good approximation for polycrystalline solids and mixtures.
The compressional and shear sound velocities are
\be
v_P = \lk \frac {K+ \frac 43 G}{\rho}\rk^{\frac 12}, \quad
v_S = \lk \frac {G}{\rho}\rk^{\frac 12}.
\ee
This is often written as $v_P^2 = v_B^2 + \frac 43 v_S^2$ with
$v_B = \lk {K}/{\rho}\rk^{\frac 12}$ being the bulk (hydrodynamic)
sound velocity. It has no direct physical meaning for a crystal, however
it corresponds to the sound velocity of the liquid phase, as can be seen
at the liquid-solid transition in Ref. \onlinecite{zha04prb70:174107}.
The Poisson's ratio is defined as:
\begin{equation}
\label{poisson-def}
\sigma \equiv \frac{1}{2} \frac{3K-2G}{3K+G} = 
\frac 12 \frac{3v_B^2-2v_S^2}{3v_B^2+v_S^2} =
\frac 12 \frac{v_P^2 - 2 v_S^2}{v_P^2 - v_S^2}.
\end{equation}

In order to find the values of $K$ and $G$ consider a polycrystalline solid
consisting of grains of all possible orientation, and use Voigt and
Reuss estimates of its elastic moduli. 
In the Voigt approach a uniform \emph{strain} field $u_{ij}$ (of either uniform compression or a
shear type) is applied to the mixture, and $K$ and $G$ are found from the averaged elastic
energy density. For a polycrystal this means averaging the elastic energy over all possible
orientations of the crystal relative to the strain field $u_{ij}$.
Reuss approach uses uniform \emph{stress} instead.
The bulk moduli defined in this way are equal to $K_V$ and $K_R$ defined above.
For a hexagonal lattice the bulk and shear moduli are \cite{watt80jap51:1525}
\be
K_V = \frac 19 \lk  2 C_{11} + C_{33} + 2 C_{12} + 4 C_{13} \rk, \;
K_R = \frac {Q}{C_S},
\ee
\be
G_V = \frac {1}{30} \lk C_S + 12 C_{44} + 12 C_{66} \rk,
\ee
\be
G_R = \frac 52 \; \frac {Q C_{44} C_{66}} {3 K_V C_{44} C_{66} + Q (C_{44} + C_{66})}.
\ee
Finally we use Voigt-Reuss-Hill average scheme to obtain $K$ and $G$: 
\be
K=\frac{K_V+K_R}{2}, \quad
G=\frac{G_V+G_R}{2}.
\ee

\subsection{Zero-point vibrations}
In our calculations we have treated zero-point vibrations 
within the framework of the Debye model. As discussed above,
this approach is by no means exact, especially at
low pressures, but it gives us a good estimate of the effect of ZPV
on the elastic properties. Using $K$ and $v_S$ obtained
in the absence of ZPV, we calculate the Debye temperature
\be
T_D = \frac{\hbar}{k_B} \lkv \frac{V_{mol}}{18 \pi^2 N_A} 
\lk \frac{1}{v_P^3} + \frac{2}{v_S^3} \rk \rkv^{-1/3},
\ee
zero-point energy
\be
E_{zp} = \frac 98 N k_B T_D,
\ee
and the corresponding contributions to pressure and the bulk modulus:
\be
\triangle p =  - \frac{d E_{zp}}{dV} = - \frac 98 N_A k_B \frac {d T_D}{d V_{mol}},
\ee
\be
\triangle K =  - V_{mol} \frac{d \triangle p}{dV_{mol}} =
  \frac 98 V_{mol} N_A k_B \frac {d^2 T_D}{d V_{mol}^2}.
\ee
We find the volume derivatives of $T_D$ analytically, using the
Rose-Vinet expression for $K$ and a simple parametrization for $v_S$.
Below we always compare results with and without ZPV in order to measure
the importance of quantum effects on each physical quantity.
$^4$He isotope was assumed for all our ZPV calculations.
\section{\label{s:results}Results and discussion}
\subsection{Equation of state, metallization and the $c/a$ ratio}

\begin{figure}
\includegraphics[scale=0.32]{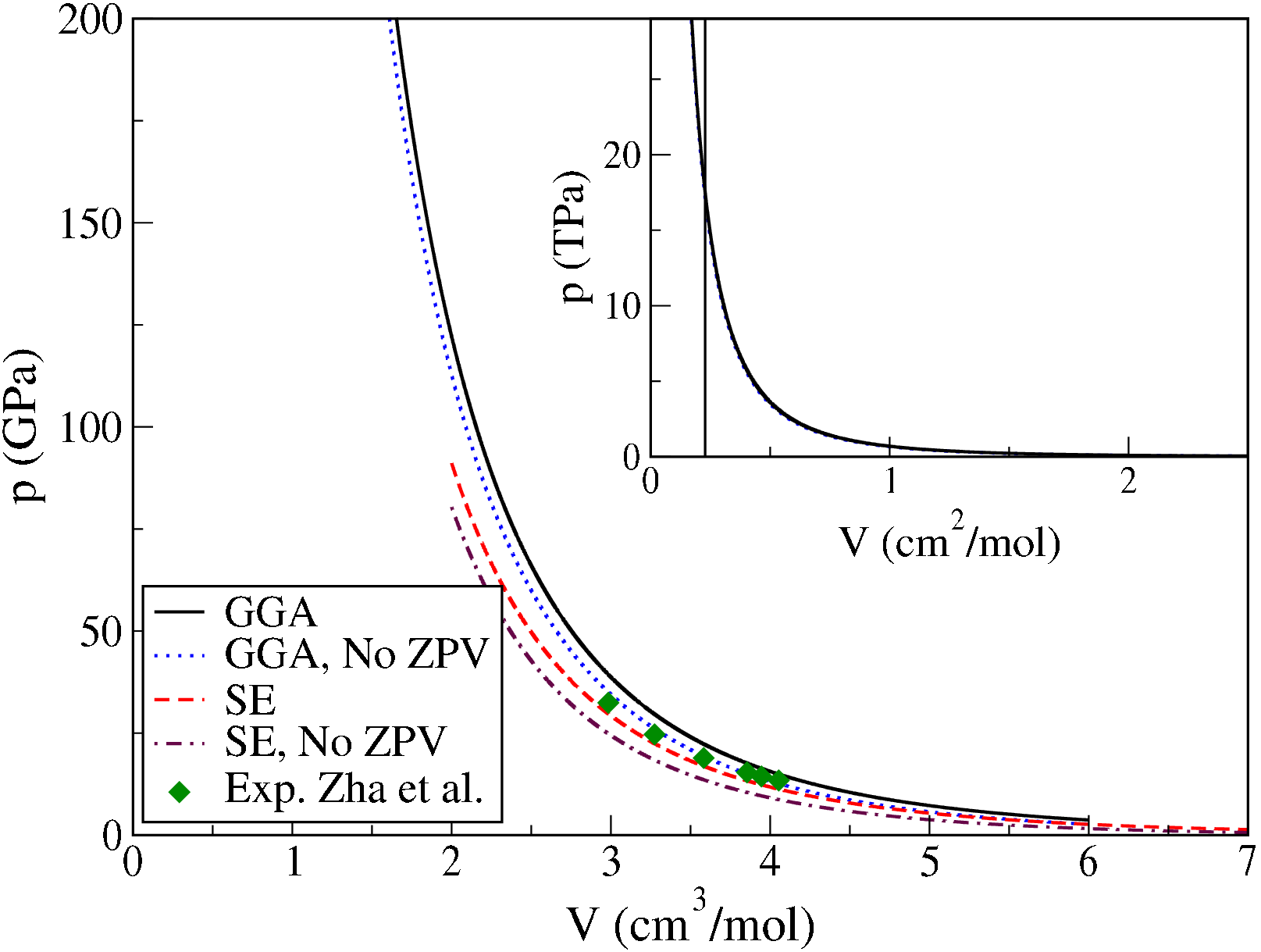}
\caption{\label{f:eos} (Color online) Equation of state of hcp helium. Inset: high-pressure region.
The vertical line indicates the GGA metallization point.}
\end{figure}

The calculated equations of state (EOS) $p(V_{mol})$
of the hcp helium are presented in Fig. \ref{f:eos}
and compared with the experimental data \cite{zha04prb70:174107}.
The four curves correspond to our four approaches: SE and GGA, with and without ZPV.
EOS can be viewed as an auxiliary quantity in our calculations, as it is
used to calculate Reuss bulk modulus. It is also used to change variables from
$V_{mol}$ to $p$, which is done to obtain $p$-dependent quantities presented
in all figures of the present paper. For this we always used
the respective EOS for each of the four approaches, e.g. a GGA+ZPV EOS 
was used for GGA+ZPV elastic moduli, but SE (No ZPV) EOS was used for SE (No ZPV) 
elastic moduli.

The semi-empirical potentials underestimate the pressure $p$ for a given $V_{mol}$,
but zero-point vibrations improve the agreement with the experiment significantly.
GGA overestimates the pressure, and the inclusion of ZPV makes things worse.
The effect of ZPV is more pronounced in the SE approach, as the SE pressure 
for a given $V_{mol}$ is smaller compared to GGA. 

The inset of Fig. \ref{f:eos} shows the high-pressure EOS (GGA with and without ZPV)
for pressures up to the metallization point and above. The metallization takes place
at $V_{mol} = 0.228$ cm$^3$/mol in 
our GGA calculations \cite{grechnev10fnt36:423}
($p= 17.48$ TPa from our GGA+ZPV EOS, or $p=17.08$ TPa without ZPV).
This point is shown as the vertical line in the inset of Fig. \ref{f:eos}.
Our metallization volume and pressure are in good agreement with
previous GGA results \cite{khairallah08prl101:106407,monserrat14prl112:055504}.
Note, however, that GGA seriously 
overestimates the metallization volume (by about 20$\%$ for He according
to the diffusion Monte-Carlo and GW studies \cite{khairallah08prl101:106407,monserrat14prl112:055504}),
thus underestimating the metallization pressure. Moreover, it has been recently
shown \cite{monserrat14prl112:055504} that vibrational degrees of freedom further
increase the metallization pressure, but these effect cannot be reproduced in harmonic approximation.
Ref.  \onlinecite{monserrat14prl112:055504} gives the value $p=32.9$ TPa.
Such questions are beyound the scope of the present paper. We use only GGA with 
ZPV in harmonic Debye approximation, which results in an underestimated metallization pressure of
$17.48$ TPa. 

Our calculated $c/a$ ratios have been presented previously in
Refs. \onlinecite{freiman09prb80:094112,grechnev10fnt36:423}. SE and GGA give somewhat
different $c/a(V_{mol})$ curves, but both methods give lattice distortions
$\delta \equiv c/a - \sqrt{8/3}$ of the order of $10^{-3}$ at the pressures
up to 150 GPa. $\delta$ is negative for $p > 13$ GPa in both approaches.
For the TPa pressures the magnitude of the negative $\delta$ obtained with GGA grows
and it reaches a sharp minimum $\delta \approx -0.05$
at the metallization point $V_{mol}=0.228$ cm$^3$/mol (Ref. \onlinecite{grechnev10fnt36:423}).
\subsection{Elastic moduli and the elastic anisotropy}

\begin{figure}
\includegraphics[scale=0.38]{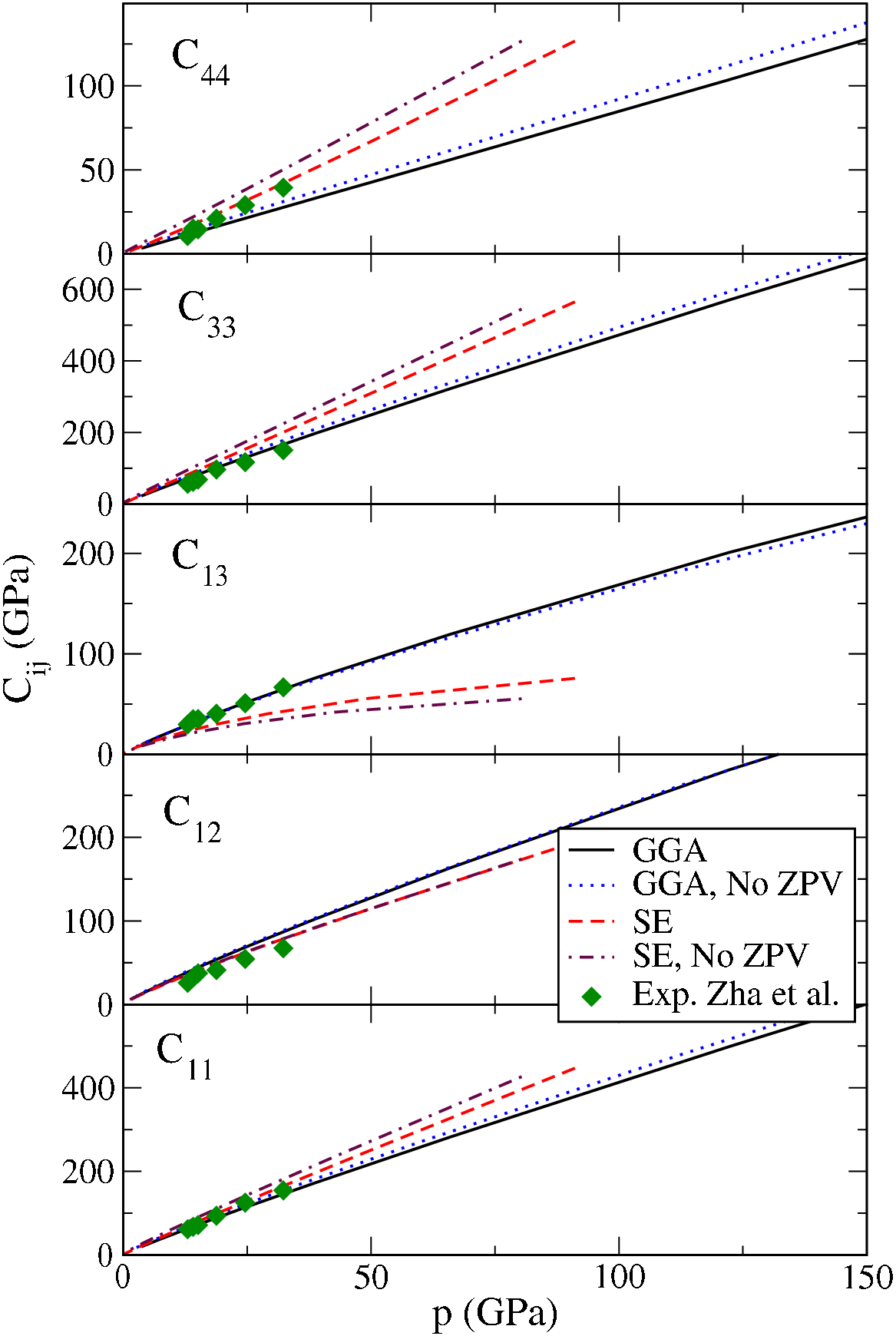}
\caption{\label{f:elastic} (Color online) Five elastic constants of hcp helium versus pressure.}
\end{figure}

The five elastic moduli for the pressures up to 150 GPa are presented in 
Fig. \ref{f:elastic}. Again, we compare our four theoretical 
approaches with the experimental data of Zha \emph{et al.} \cite{zha04prb70:174107}. 
Note that all our calculations were done for $T=0$, while the experimental
elastic moduli were measured at room temperature $T=300$ K, which can be one of the
reasons for the theory-experiment discrepancies. Both SE and GGA
are in reasonable agreement with the experiment, although neither method
gives a perfect quantitative match. GGA seems to be more
consistent of the two.
Our GGA results without ZPV are close to the GGA and LAG results obtained
in Ref. \onlinecite{nabi05prb72:172102}. 
The DFT-GGA errors mainly stem from the inaccuracy of the GGA functional itself,
we have checked that the parameters of the FP-LMTO calculations (number of $k$-points,
tail energies, etc.) have minimal effect on the results.
It is unlikely that the errors of the SE approach (at least for the pressures
$p \lesssim 50$ GPa) are caused solely by the neglect 
of the $4$-body and higher order $n$-body forces. The most likely reason
for the SE-experiment differences (apart from the temperature effect) is that
the elastic moduli are sensitive to the particular parametrization
\cite{freiman07fnt33:719} of the pair and triple forces.
The effect of ZPV, while noticeable, does not affect the elastic moduli 
in any drastic way.

\begin{figure}
\includegraphics[scale=0.3]{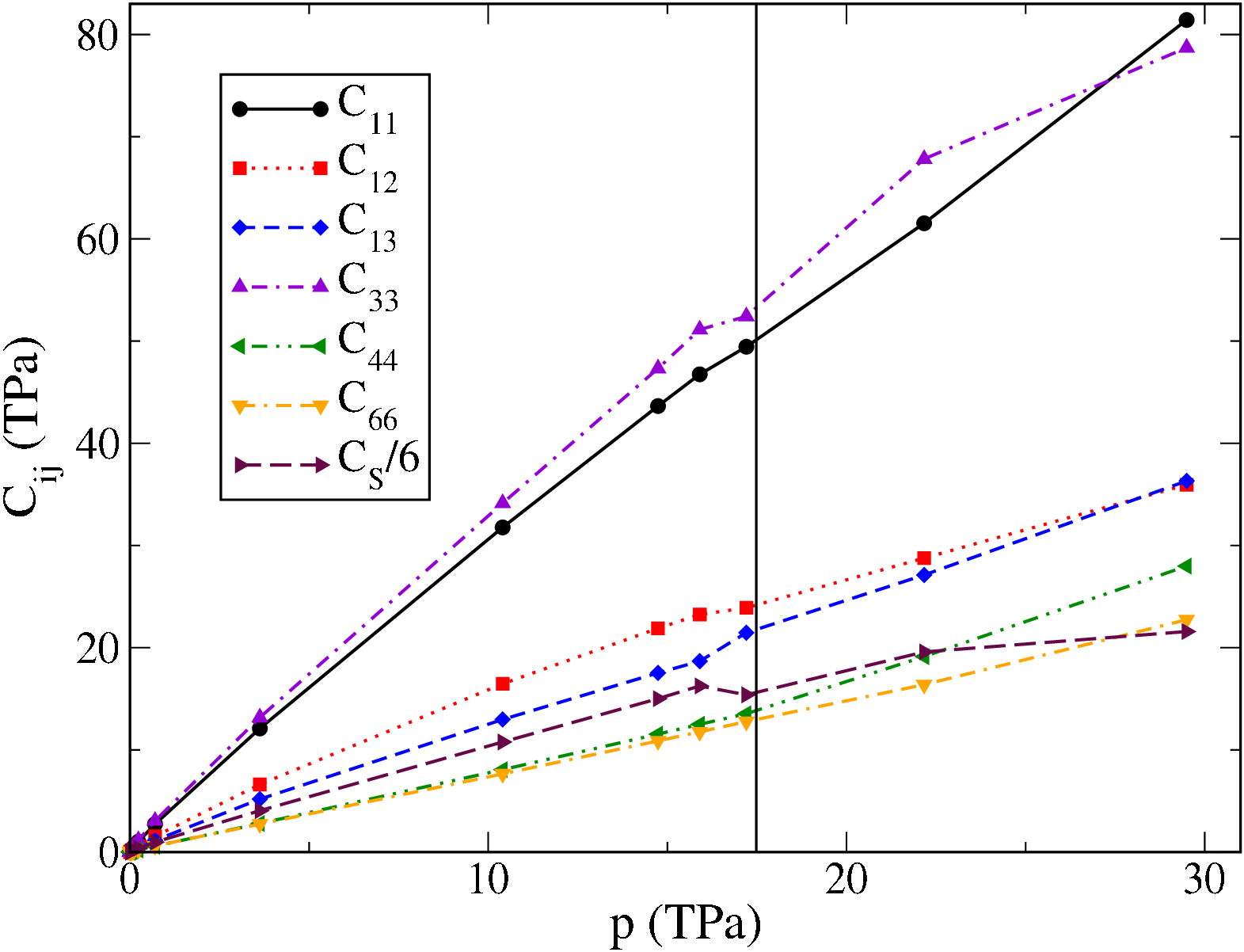}
\caption{\label{f:elastic-high} (Color online) Elastic moduli (GGA+ZPV only) of hcp helium
under terapascal pressures. Symbols are the calculated data points.
The vertical line indicates the GGA metallization point. 
$C_{66} \equiv (C_{11}-C_{12})/2$ and $C_S/6$ are also presented.} 
\end{figure}

In Fig. \ref{f:elastic-high} the calculated elastic moduli (GGA+ZPV) are
presented for the terapascal pressure range. The two additional shear moduli
$C_{66}$ and $C_S/6$ are also shown. Significant features  are seen at 
the GGA metallization point $p=17.48$ TPa. They are most likely finite
discontinuities (steps), however, critical behavior at the metallization transition
is beyond the scope of the present paper and would be difficult to investigate accurately
with the methods we employ.  Of the three elemental shear 
moduli only $C_S/6$ has a large jump ($\approx -5 \%$) at the metallization point,
while the behavior of $C_{44}$ and $C_{66}$ 
is relatively smooth. Remember that $C_S$ is the measure of stiffness of the crystal
with respect to the isochoric change of the $c/a$ ratio, so it is not unexpected 
that the behavior of $C_S$ is irregular at the point where 
$c/a$ has a sharp minimum \cite{grechnev10fnt36:423}. 
Note that $C_S(p)$ also behaves rather nonlinearly in the metallic phase.
These irregularities in $C_S$ affect the five elastic constants, most notably
$C_{13}$ and $C_{33}$, via Eqs. (\ref{modulus1})--(\ref{modulus3}).
Hcp lattice is dynamically stable for the whole pressure range considered,
i.e. the conditions \cite{nabi05prb72:172102} $C_{44}>0$, $C_{11} > \lab C_{12}\rab$,
$C_{11} C_{33} > (C_{13})^2$ and $C_{33} (C_{11} + C_{12}) > 2 C_{13}^2$ are
fulfilled, which ensures that the single-crystal sound velocities
(\ref{sc-vel1})--(\ref{sc-vel3}) are real.

All our calculations were done for the relaxed $c/a$ ratio,
and the effect of the volume dependence of $c/a$ has been taken into account
through the parameter $R$. In order to test the importance of the $c/a$ distortion
for the elastic moduli we have also performed calculations with $R=0$, like
in Ref. \onlinecite{nabi05prb72:172102}. Our results (not shown) indicate that
the difference is barely visible for the pressures up to 150 GPa, however,
it becomes significant (about 5\%) for the TPa pressure range, and it noticeably
affects the features near the metallization point. We took great care in
choosing the parametrization of the function $c/a(V_{mol})$  that
reproduces the sharp minimum in $c/a$ well.

\begin{figure}
\includegraphics[scale=0.35]{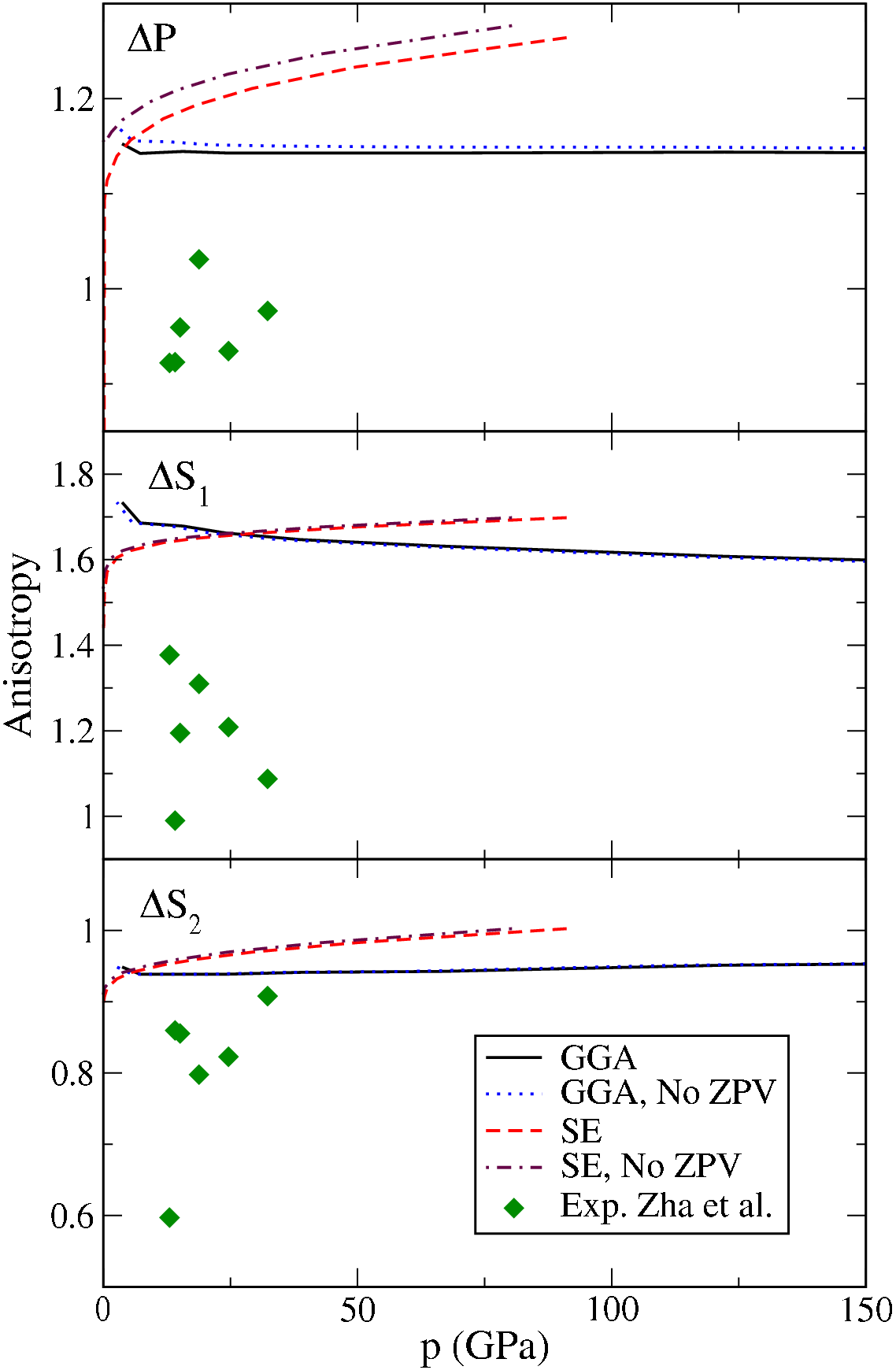}
\caption{\label{f:anis} (Color online) Elastic anisotropy parameters of hcp helium versus pressure.}
\end{figure}

The elastic anisotropies $\triangle P$, $\triangle S_1$ and $\triangle S_2$
are presented in Fig. \ref{f:anis}. The compressional anisotropy $\triangle P$ 
is close to the isotropic value of $1$  in the experiment, however both GGA and SE
predict noticeable anisotropy $\triangle P > 1$. GGA gives a virtually pressure independent
value $\triangle P \approx 1.14$, while the semi-empirical $\triangle P$ grows 
from $\approx 1.1$ at low pressures to $\approx 1.25$ at 100 GPa. 
$\triangle S_1$ is about $1.2-1.3$ in the experiment, both GGA and SE overestimate
it significantly, giving values of the order of $1.6-1.7$. The third parameter, 
$\triangle S_2$, is always less then one in both theory and experiment.
SE and GGA, however,  give values closer to $1$ than the experiment,
thus underestimating the anisotropy. 

\begin{figure}
\includegraphics[scale=0.4]{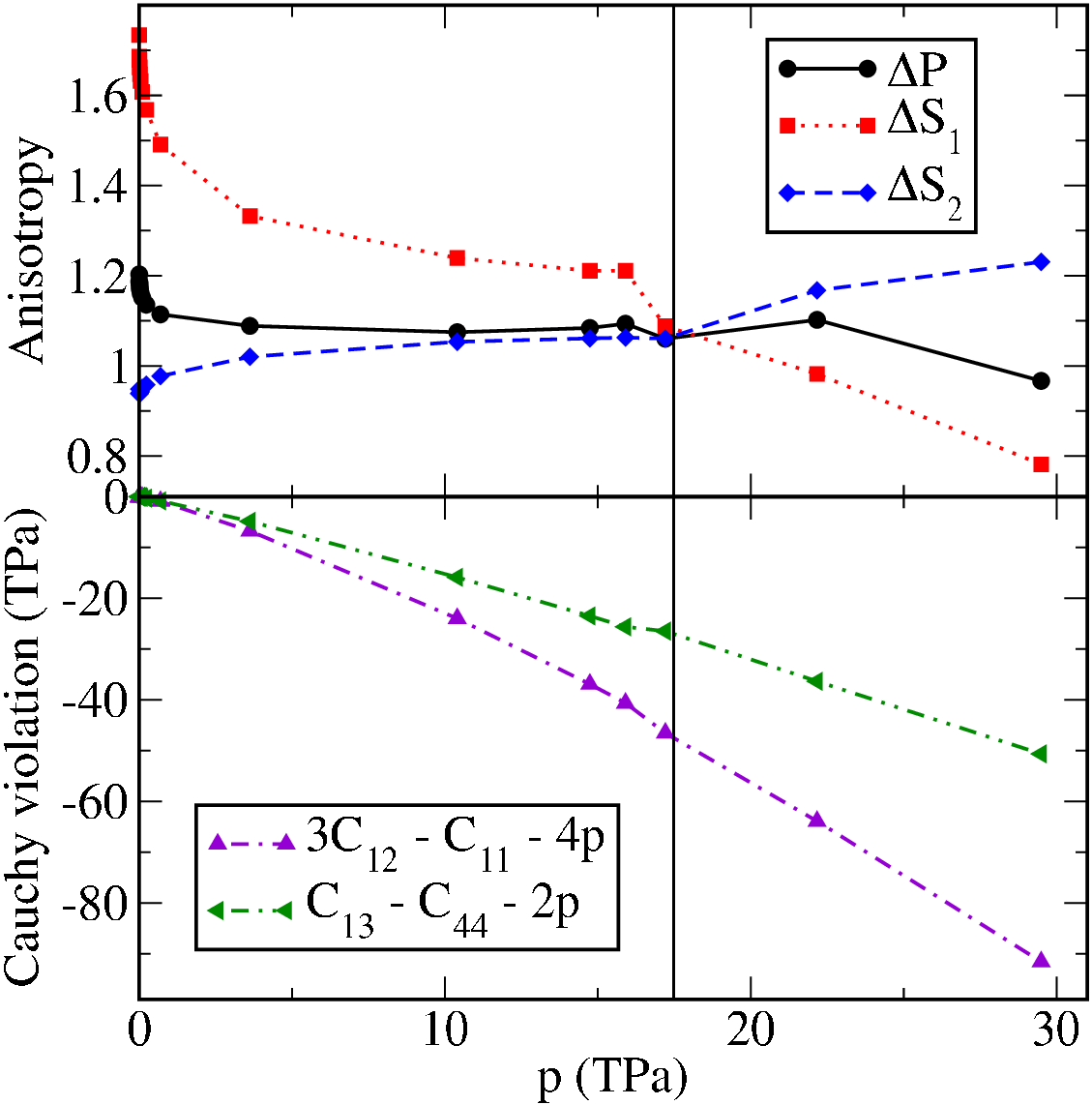}
\caption{\label{f:anis-high} (Color online) Elastic anisotropy parameters (upper panel) and
Cauchy violations (lower panel) of hcp helium under terapascal pressures.
Symbols are the calculated data points.
The vertical line indicates the GGA metallization point. }
\end{figure}

To summarize, GGA gives $\triangle P \approx 1.14$, $\triangle S_1 \approx 1.7$,
$\triangle S_2 \approx 0.93$ and the pressure dependences of these parameters are rather small.
SE approach, while agreeing well with GGA for low pressures,
predicts much stronger pressure dependence of $\triangle P$, $\triangle S_{1,2}$, and different
sign of $d \triangle S_1/d p$. 
Ref. \onlinecite{nabi05prb72:172102} states that the three anisotropy parameters are 
nearly pressure-independent. While our GGA data fully confirms this result,
our SE data behaves quite differently.
In the absence of reliable experimental data on the pressure
dependences it is hard to say which behavior is more correct (see the discussion below, however).
Both methods overestimate $\triangle P$ and $\triangle S_1$ significantly, and also overestimate
$\triangle S_2$  (which underestimates the anisotropy), thus
neither approach agrees particularly well with the experiment.
In particular, the relative errors in $\triangle P$, $\triangle S_1$ and $\triangle S_2$
(computed relative to the experiment) are significantly larger than the errors in $C_{ij}$.
The reasons for such discrepancy are unknown. Since both GGA and SE agree with each other better
than with the experiment, one can speculate that the difference in temperature between
0 K and 300 K might play at least some role. The large elastic anisotropy of He,
while somewhat unexpected,
is not in any way incompatible with the high symmetry of the hcp crystal
and the nearly spherical shape of the atoms. Highly symmetric crystals
have isotropic or nearly isotropic rank-2 tensor properties, such as conductivity,
thermal expansion or
dielectric permittivity. The stiffness, however, is a rank-4 tensor property, 
and it is well-known that even cubic crystals are not elastically isotropic.
The large values of $\triangle S_1$ are not very surprising, since
this parameter is similar to the single anisotropy parameter $(C_{11}-C_{12})/(2C_{44})$
of the cubic crystal.

The three anisotropy parameters at TPa pressures are presented in Fig. \ref{f:anis-high}, upper panel.
In the insulator phase anisotropy decreases with pressure, with three parameters becoming close
to unity just before the metallization transition. This is an intuitively plausible
behavior of atomic crystal becoming more isotropic under pressure.
However, $\triangle P$, $\triangle S_1$ and $\triangle S_2$ demonstrate  strong features
at the metallization point (especially $\triangle S_1$, which involves $C_S$),
with anisotropies of two shear modes changing sign at the metallization transition, 
and $\triangle P$ staying close to one and showing
nonlinear behavior in the metallic phase. One can understand the anisotropy parameters in terms of 
the elastic moduli presented in Fig. \ref{f:elastic-high}. For instance, let us analyze $\triangle S_1$.
For an isotropic solid 
$C_{44} = C_{66}= C_S/6$. For helium at low pressures, however, $C_S/6$ has almost twice the value
of $C_{44}$, giving a large $\triangle S_1$. At higher pressures $C_S/6$ is of
the same order as $C_{44}$ and $C_{66}$, and eventually becomes smaller than them in 
the metallic phase, which corresponds to $\triangle S_1 < 1$.

\begin{figure}
\includegraphics[scale=0.4]{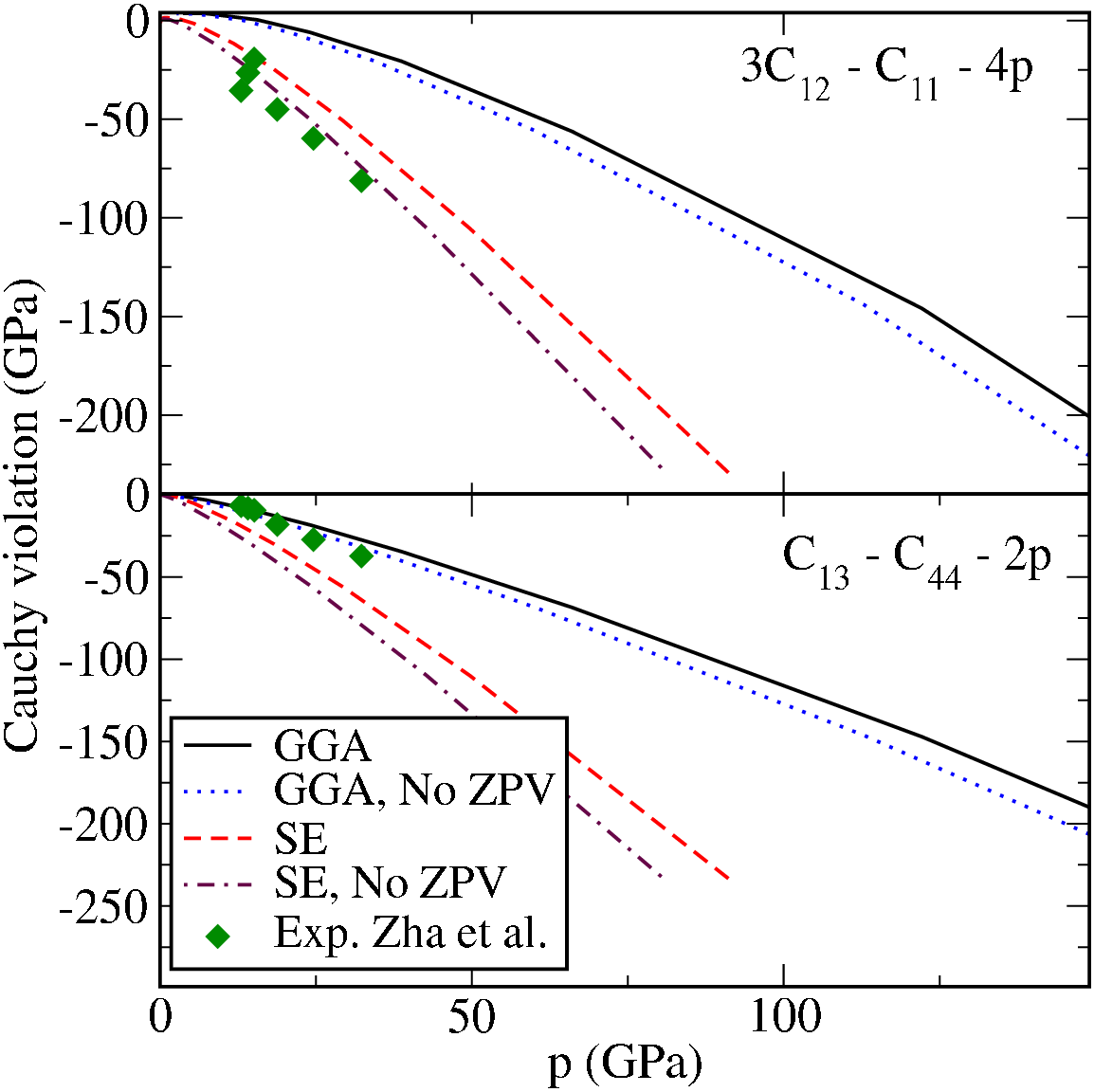}
\caption{\label{f:cauchy} (Color online) Cauchy violations of hcp helium versus pressure.}
\end{figure}

Strictly speaking, it is wrong to say that the anisotropy parameters are nearly pressure-independent.
Indeed, the pressure-induced changes in Fig. \ref{f:anis-high} are rather dramatic even if we consider 
the insulating phase only. The pressure-independence found in Fig. \ref{f:anis}, and in Ref.
\onlinecite{nabi05prb72:172102}, is simply the result of the pressure scale of $150$ GPa being
very small compared to the metallization pressure $\sim 17$ TPa, which is presumably the only natural pressure 
scale in solid helium (if the quantum effects are disregarded).  From this logic we can infer 
that GGA is more trustworthy than SE in determining pressure dependence of $\triangle P, \triangle S_{1,2}$,
since it gives reasonable results in TPa range of pressures, while the SE method breaks down at
$p \sim 100$ GPa, which can act as a spurious pressure scale. This argument is far from infallible,
however, as GGA is known to be unreliable for pressures $p \lesssim 50$ GPa due to the
poor description of the van der Walls forces.

\begin{figure}
\includegraphics[scale=0.35]{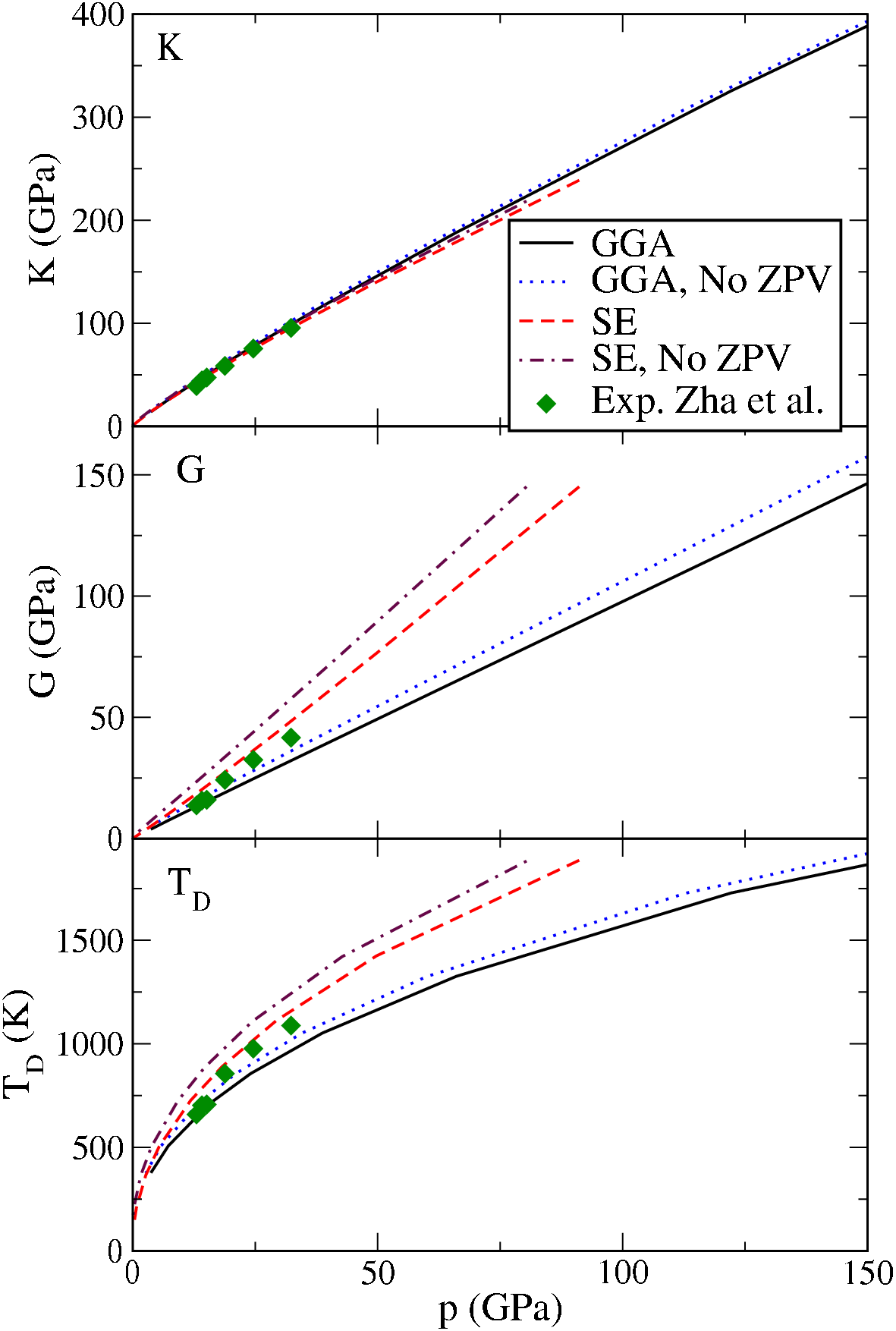}
\caption{\label{f:aggr} (Color online) Aggregate properties of hcp helium versus pressure:
 bulk modulus $K$ (upper panel), shear modulus $G$ (middle panel) and Debye temperature (lower panel).
 $K$ and $G$ are obtained using Voigt-Reuss-Hill averaging.}
\end{figure}

The Cauchy violations $3 C_{12} - C_{11} - 4 p$ and $C_{13} - C_{44} - 2 p$ are presented
in Fig. \ref{f:cauchy}.  They can be thought of as a measure of noncentral forces in a solid. 
Experimental $3 C_{12} - C_{11} - 4 p$ is reproduced well by SE and not so well by GGA,
but the situation is reversed for $C_{13} - C_{44} - 2 p$. The Cauchy violations
at terapascal pressures are shown in Fig. \ref{f:anis-high}, lower panel. Their behavior
is mostly linear with mild kinks at the metallization point.

\subsection{Aggregate properties}

\begin{figure}
\includegraphics[scale=0.4]{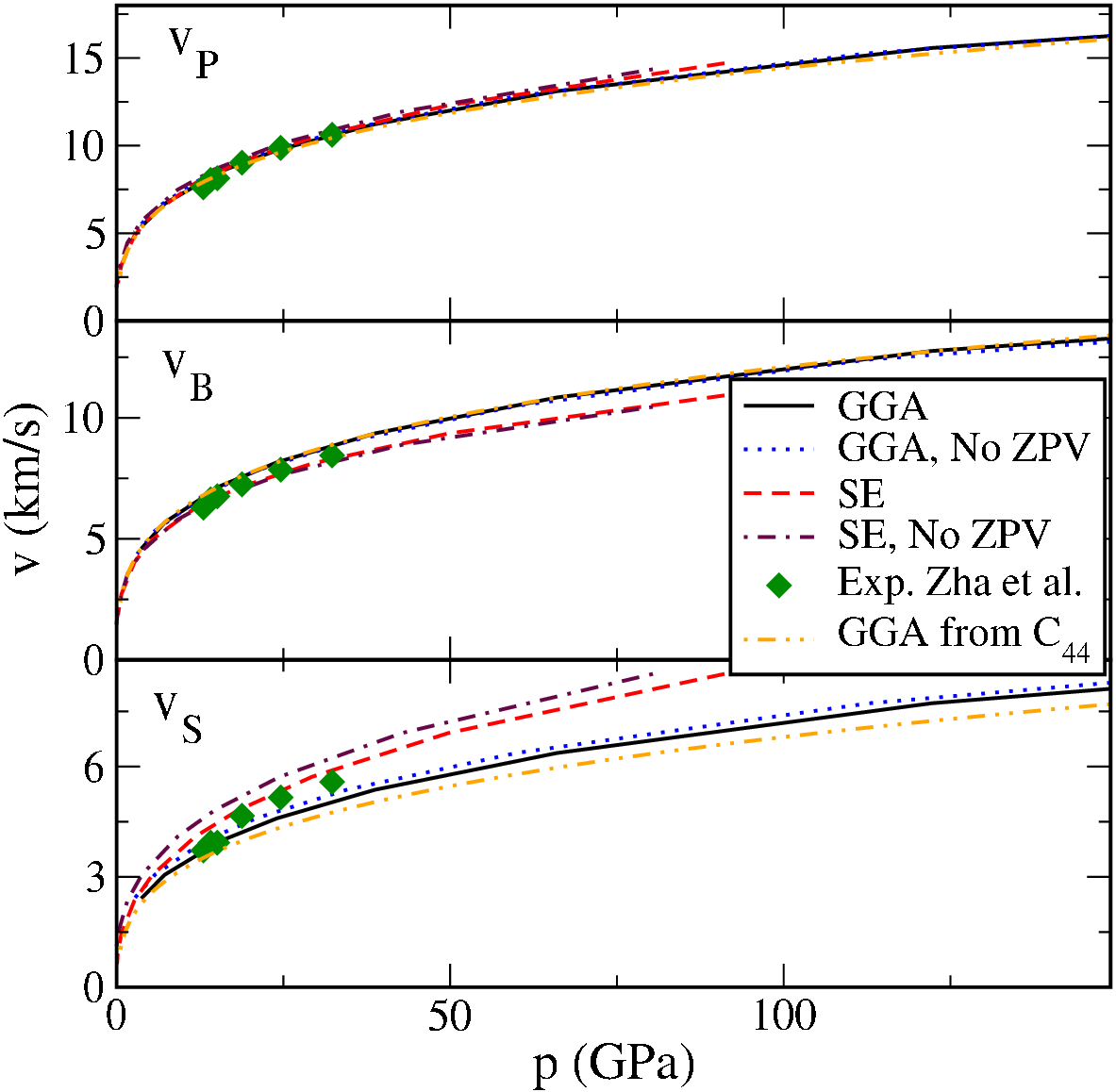}
\caption{\label{f:soundvel} (Color online) Aggregate sound velocities of hcp helium versus pressure.}
\end{figure}

The Voigt-Reuss-Hill-averaged bulk and shear moduli are presented in Fig. \ref{f:aggr}.
For the bulk modulus $K$ both methods agree very well with the experiment and the difference
between GGA and SE is small. For the shear modulus $G$, however, there is a significant
difference between GGA and SE. Fig. \ref{f:soundvel} shows the three sound velocities
$v_P$, $v_B$ and $v_S$. Both GGA and SE agree with the experiment pretty well, with
$v_S$ (which is proportional to $\sqrt{G}$) displaying the largest SE-GGA difference.
We also plot $\sqrt{C_{44}/\rho}$
(calculated with GGA+ZPV) as the golden dash-dot-dot curve in the lower panel of
Fig. \ref{f:soundvel}. It has a meaning of $v_S$ calculated by using the modulus $C_{44}$
instead of the averaged shear modulus $G$, like we did previously in
Ref. \onlinecite{freiman13prb88:214501}.
For an isotropic solid $G=C_{44}$. For helium, using $C_{44}$
instead of $G$ underestimates $v_S$ by a few percent and worsens the agreement between
GGA and the experiment. In other words, by calculating $v_S$ 
from the proper Voigt-Reuss-Hill $G$ in the present work
we actually improve the GGA-experiment agreement compared to
Ref. \onlinecite{freiman13prb88:214501}.

\begin{figure}
\includegraphics[scale=0.4]{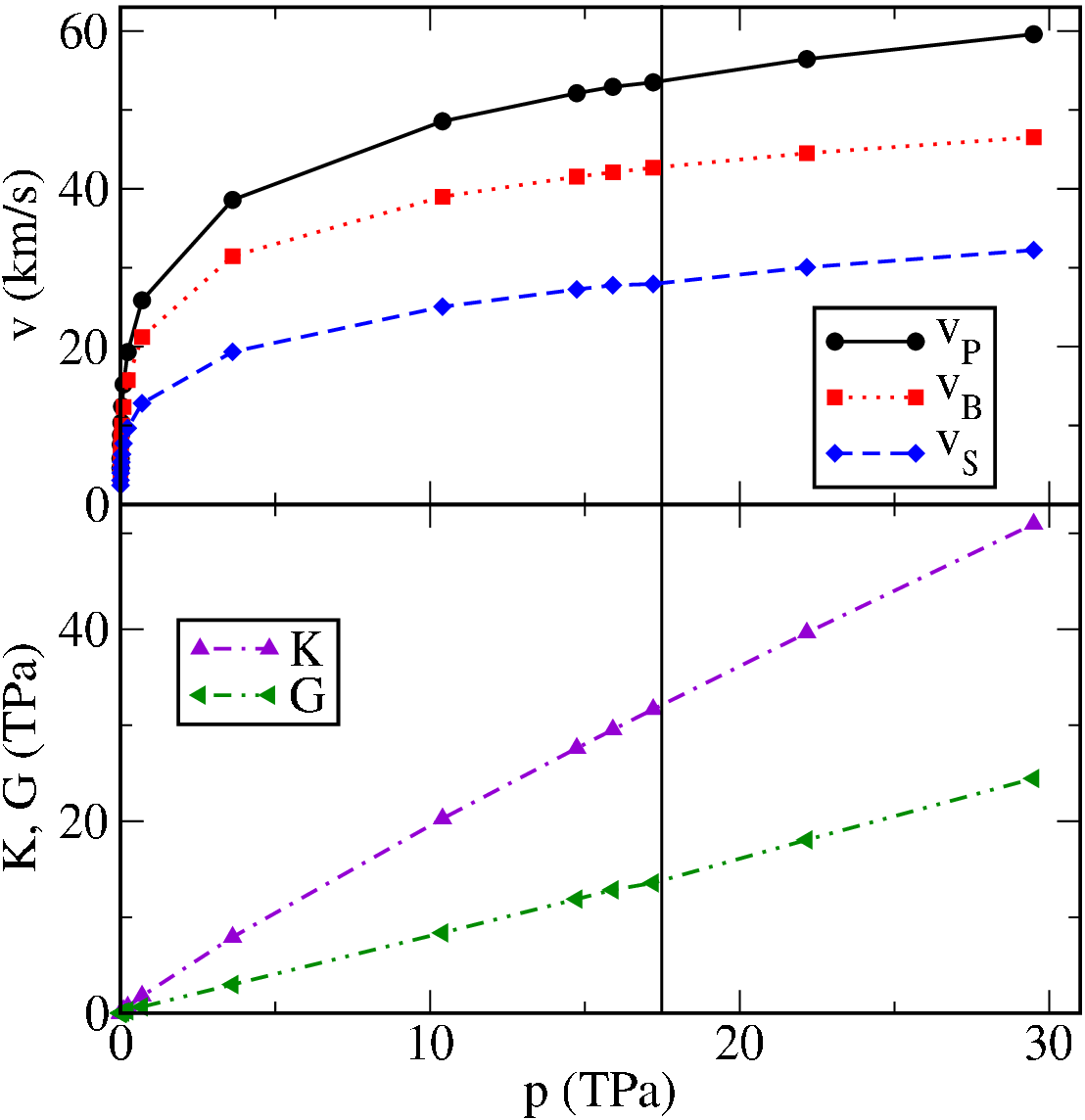}
\caption{\label{f:aggr-high} (Color online) Aggregate properties of hcp helium 
in the TPa pressure range: sound velocities (upper panel), bulk and shear moduli (lower panel).
Symbols are the calculated data points.
The vertical line indicates the GGA metallization point. }
\end{figure}

The bulk and shear moduli and sound velocities at TPa pressures are presented in
Fig. \ref{f:aggr-high}. All these quantities show features at the metallization point,
with kinks in $G$ and, respectively, $v_S$, being the most pronounced. 

The Debye temperature $T_D$ is presented in Fig. \ref{f:aggr}, lower panel. Just like with 
shear modulus $G$, there is a significant difference between SE and GGA, with experimental
data points lying in the middle.  The Debye temperature for the TPa pressures is
plotted in Fig. \ref{f:poisson-high}, upper panel. It shows a noticeable feature at the 
metallization point.

\subsection{Poisson's ratio}

\begin{figure}
\includegraphics[scale=0.32]{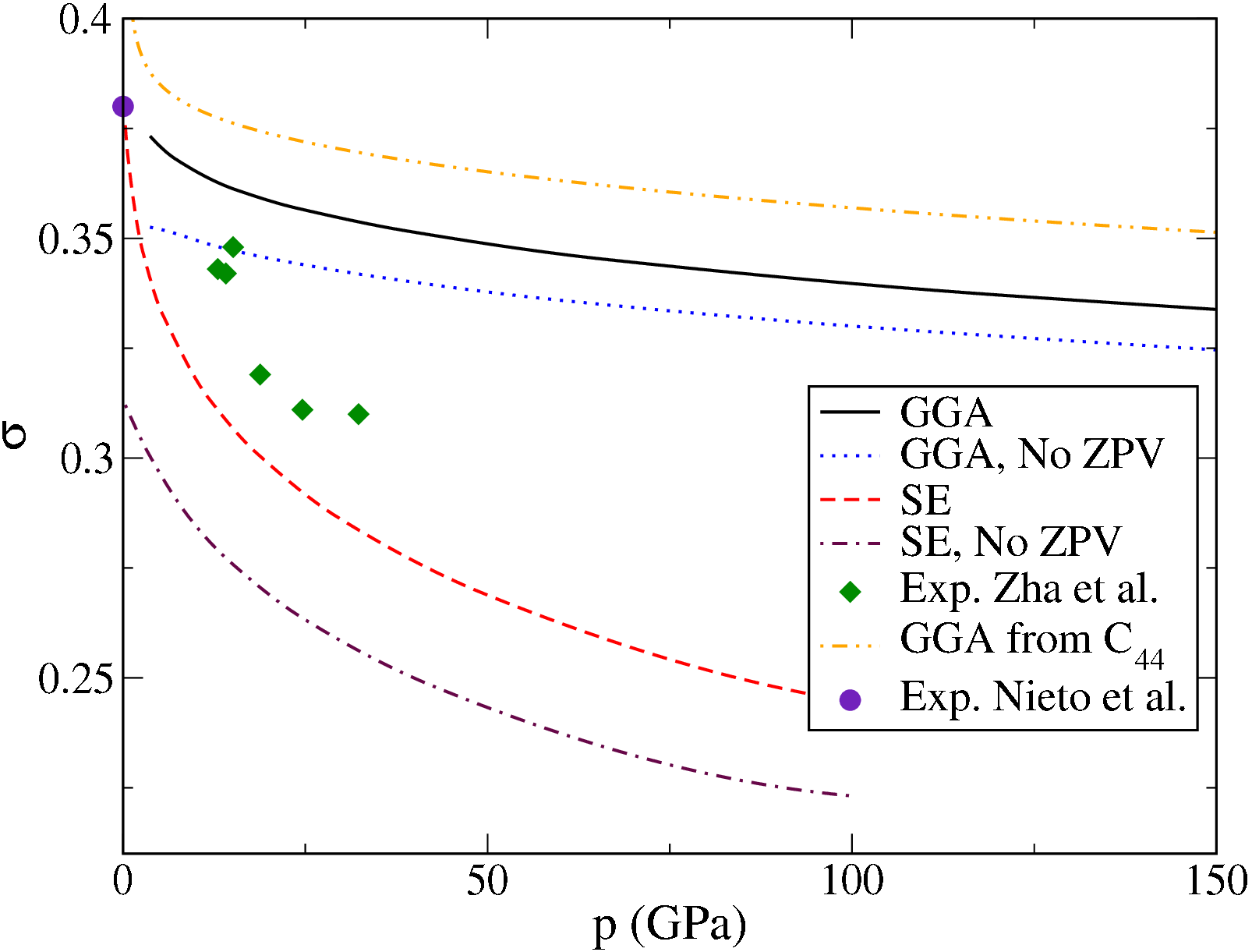}
\caption{\label{f:poisson} (Color online) Poisson's ratio of hcp helium versus pressure.}
\end{figure}

Figure \ref{f:poisson} shows the Poisson's ratio $\sigma$ as a function of pressure. 
The GGA+ZPV results \cite{freiman15fnt41:571},
obtained using $C_{44}$ instead of $G$, as explained above, are also plotted
as the golden dash-dot-dot curve. In addition to the experimental data of Zha. \emph{et al.}
\cite{zha04prb70:174107}, we also plot the low-$p$, low-$T$ data of Nieto  \emph{et al.}\cite{nieto12njp14:013007}.
Since the Poisson's ratio (\ref{poisson-def}) is
proportional to the difference $3K - 2G$, it is a rather method-sensitive quantity.  
Just like for the elastic anisotropy parameters above, SE gives much
stronger pressure dependence of $\sigma$ compared to GGA, and for the reasons outlined above
for $\triangle P, \triangle  S_{1,2}$ we find GGA results more trustworthy in this aspect. 
On the other hand, the experimental pressure dependence of $\sigma$ seems to be closer
to the SE one.
Neither method agrees perfectly with the $T=300$ K experimental data of Zha. \emph{et al.}
\cite{zha04prb70:174107}, however, \emph{both} our method agree \emph{quantitatively} with
the $p \approx 0, T \approx 0$ result of Nieto  \emph{et al.}\cite{nieto12njp14:013007}
($\sigma=0.38$, the purple dot in Fig.  \ref{f:poisson}) as long as ZPV is included.

\begin{figure}
\includegraphics[scale=0.4]{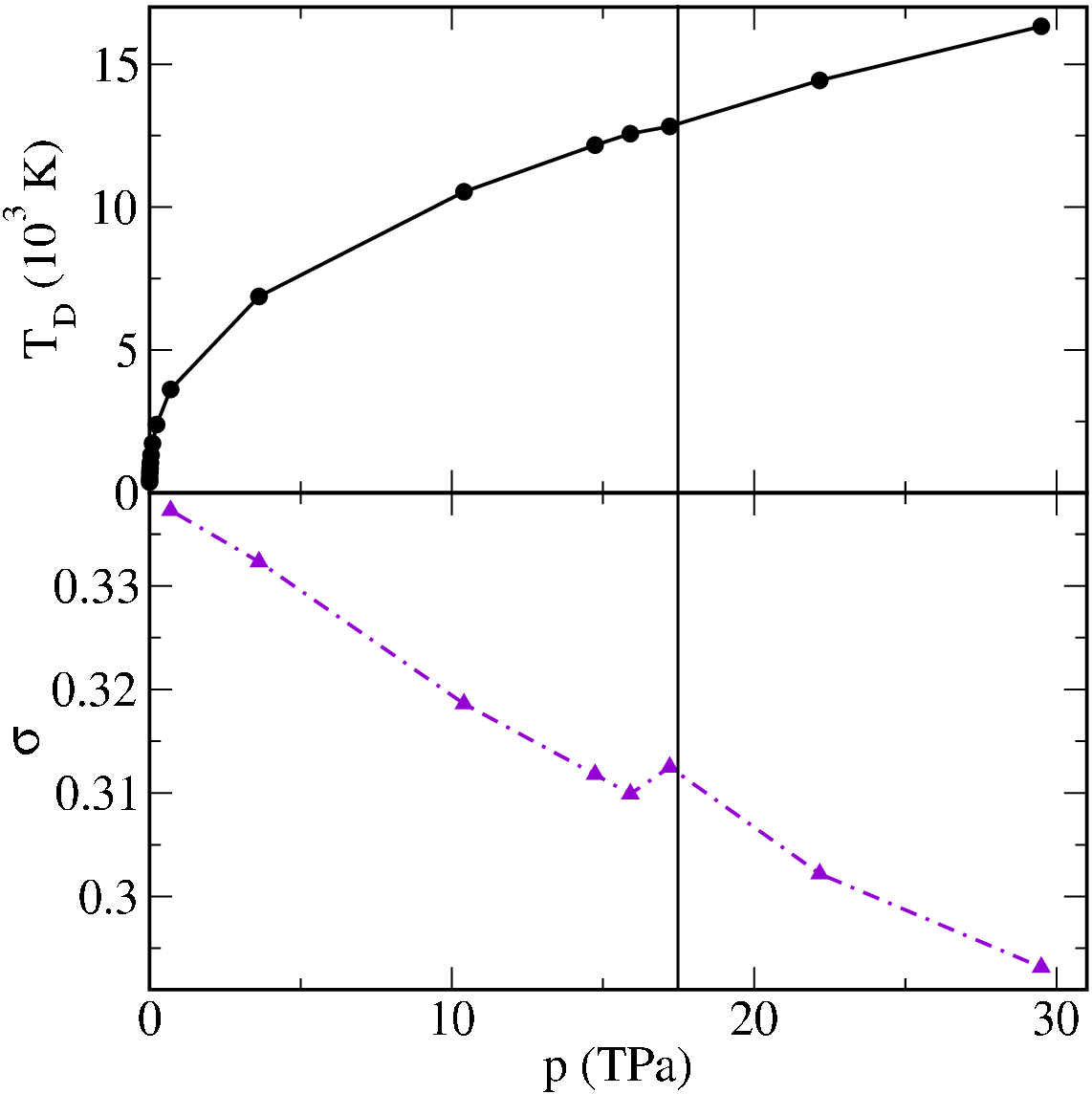}
\caption{\label{f:poisson-high} (Color online) Debye temperature (upper panel) and
Poisson's ratio (lower panel) of hcp helium  in the terapascal pressure range.
Symbols are the calculated data points.
The vertical line indicates the GGA metallization point. }
\end{figure}

Although different theoretical and experimental methods give somewhat different values
of $\sigma$, they all agree upon one fact: $d \sigma/d p <0$, i.e. $\sigma$ decreases monotonously
when pressure increases, fully confirming the surprising result of Zha. \emph{et al.}
In particular, while the role of ZPV for $\sigma$ is somewhat larger than for other 
quantities studied in the present paper, and $\sigma$ shows an isotopic effect \cite{nieto12njp14:013007},
the negative pressure dependence of $\sigma$ is definitely not a quantum effect, as
the "No ZPV" curves show the same sign and order of magnitude of $d \sigma/d p$.
This behavior of $\sigma$ is highly unusual, although solid hydrogen 
\cite{zha93prb48:9246,freiman15fnt41:571} also has negative $d \sigma/d p$ at least for
low pressures. All heavier rare-gas solids (RGSs) \cite{freiman15fnt41:571}
have positive $d \sigma/d p$.
The difference in behavior between He and heavier RGSs is, we repeat, is not a quantum effect,
i.e. it is not caused by the small mass of He atoms. Presumably the crucial factor here
is the difference of the outermost electron shells in He (1s$^2$ shell)
and heavier RGSs (ns$^2$np$^6$). The situation is similar for $c/a-\sqrt{8/3}$,
which have different sign for He and other RGS's in GGA \cite{grechnev10fnt36:423} .
In the language of SE potential, both types of behavior of $\sigma$
can be accounted for by using pair and triple forces of exactly the same functional
form \cite{freiman07fnt33:719}, but with different values of parameters.

The Poisson's ratio at terapascal pressures is plotted in Fig. \ref{f:poisson-high}, lower
panel. $\sigma$ decreases monotonously up to highest pressures apart from a large feature at
the metallization point. There is no minimum in $\sigma$ (disregarding the step at the metallization point)
and definitely no $\sigma \to 0.5$ high-pressure asymptotic that many materials have. 
At highest pressures considered $\sigma$ is of the order of $0.29$.

A question often asked is whether helium becomes a classical crystal under high pressure,
or in other words, whether the relative effect of ZPV on various physical quantities
decreases with pressure. This question is surprisingly nontrivial, as it depends on the
complicated interplay of the kinetic and potential energy of the He atoms. A recent
diffusion Monte Carlo analysis \cite{cazorla15prb91:024103} has shown that 
the kinetic to potential energy ratio $\lab E_{kin}/E_{pot}\rab$
(which can be viewed as the measure of quantumness)
does indeed decrease monotonuously with pressure, however this
decrease becomes very slow for pressures $p \gtrsim 85$ GPa. We have reached similar
conclusions in the present work, which is best seen for Poisson's ratio (fig. \ref{f:poisson}).
$\sigma$ is dimensionless, and the difference between classical and quantum $\sigma$ slowly
decreases with pressure. In particular, this difference is of the order of $0.01$ in the
$150$ GPa pressure range, but in the TPa range (not shown) it reaches the value of 
about $0.0035$.

\section{Conclusion}

We have calculated five elastic constants of hcp helium under pressure,
and various derived quantities measured by Zha. \emph{et al.}\cite{zha04prb70:174107}:
anisotropy parameters, sound velocities, Poisson's ratio etc.
We have analyzed these quantities both in the pressure range up to $150$ GPa,
where experimental data is available
and semi-empirical potentials are applicable;
and in the TPa pressure range (GGA only), where the metallization transition takes place.
Both methods (GGA and SE) are in general agreement with the experiment. Most
calculated quantities display noticeable features at the metallization point.
Zero-point vibrations do not affect elastic properties
of helium in any dramatic way for the pressures considered in the present paper
(disregarding the quantum crystal region at very low pressures).

Our calculations predict significant elastic anisotropy for hcp helium ($\triangle P \approx 1.14$,
$\triangle S_1 \approx 1.7$, $\triangle S_2 \approx 0.93$ at low pressures). Both GGA and SE
overestimate anisotropy parameters compared to the experiment, which might
be a temperature effect (experiments were carried out at $T=300$ K).
The three anisotropy parameters become more isotropic (close to one)
under TPa pressures, with the anisotropy of the shear modes $S_1$ and $S_2$ changing sign
at the metallization point.
Our calculated Poisson's ratio (PR) is in excellent agreement with the $T=0$, $p=0$ result
of Ref. \onlinecite{nieto12njp14:013007} ($\sigma=0.38$). 
Our calculations agree with the experimentally observed
\cite{zha04prb70:174107} decrease of the PR with pressure. 
Under TPa pressures PR reaches values $\sim 0.31$
at the metallization point ($p \approx 17.5$ TPa) and $\sim 0.29$ at $p = 30$ TPa. 
We have shown that the negative sign of the pressure dependence of PR is
not a quantum effect by performing calculations without zero-point vibrations, which yield
similar results.

\acknowledgments

We gratefully acknowledge  J. Peter Toennies for valuable discussions.

\begin{thebibliography}{26}
\expandafter\ifx\csname natexlab\endcsname\relax\def\natexlab#1{#1}\fi
\expandafter\ifx\csname bibnamefont\endcsname\relax
  \def\bibnamefont#1{#1}\fi
\expandafter\ifx\csname bibfnamefont\endcsname\relax
  \def\bibfnamefont#1{#1}\fi
\expandafter\ifx\csname citenamefont\endcsname\relax
  \def\citenamefont#1{#1}\fi
\expandafter\ifx\csname url\endcsname\relax
  \def\url#1{\texttt{#1}}\fi
\expandafter\ifx\csname urlprefix\endcsname\relax\def\urlprefix{URL }\fi
\providecommand{\bibinfo}[2]{#2}
\providecommand{\eprint}[2][]{\url{#2}}

\bibitem[{\citenamefont{Loubeyre et~al.}(1993)\citenamefont{Loubeyre,
  LeToullec, Pinceaux, Mao, Hu, and Hemley}}]{loubeyre93prl71:2272}
\bibinfo{author}{\bibfnamefont{P.}~\bibnamefont{Loubeyre}},
  \bibinfo{author}{\bibfnamefont{R.}~\bibnamefont{LeToullec}},
  \bibinfo{author}{\bibfnamefont{J.~P.} \bibnamefont{Pinceaux}},
  \bibinfo{author}{\bibfnamefont{H.~K.} \bibnamefont{Mao}},
  \bibinfo{author}{\bibfnamefont{J.}~\bibnamefont{Hu}}, \bibnamefont{and}
  \bibinfo{author}{\bibfnamefont{R.~J.} \bibnamefont{Hemley}},
  \bibinfo{journal}{Phys. Rev. Lett.} \textbf{\bibinfo{volume}{71}},
  \bibinfo{pages}{2272} (\bibinfo{year}{1993}).

\bibitem[{\citenamefont{Freiman et~al.}(2009)\citenamefont{Freiman, Tretyak,
  Grechnev, Goncharov, Tse, Errandonea, Mao, and
  Hemley}}]{freiman09prb80:094112}
\bibinfo{author}{\bibfnamefont{Yu.~A.} \bibnamefont{Freiman}},
  \bibinfo{author}{\bibfnamefont{S.~M.} \bibnamefont{Tretyak}},
  \bibinfo{author}{\bibfnamefont{A.}~\bibnamefont{Grechnev}},
  \bibinfo{author}{\bibfnamefont{A.~F.} \bibnamefont{Goncharov}},
  \bibinfo{author}{\bibfnamefont{J.~S.} \bibnamefont{Tse}},
  \bibinfo{author}{\bibfnamefont{D.}~\bibnamefont{Errandonea}},
  \bibinfo{author}{\bibfnamefont{H.-k.} \bibnamefont{Mao}}, \bibnamefont{and}
  \bibinfo{author}{\bibfnamefont{R.~J.} \bibnamefont{Hemley}},
  \bibinfo{journal}{Phys. Rev. B} \textbf{\bibinfo{volume}{80}},
  \bibinfo{pages}{094112} (\bibinfo{year}{2009}).

\bibitem[{\citenamefont{Grechnev et~al.}(2010)\citenamefont{Grechnev, Tretyak,
  and Freiman}}]{grechnev10fnt36:423}
\bibinfo{author}{\bibfnamefont{A.}~\bibnamefont{Grechnev}},
  \bibinfo{author}{\bibfnamefont{S.~M.} \bibnamefont{Tretyak}},
  \bibnamefont{and} \bibinfo{author}{\bibfnamefont{Yu.~A.}
  \bibnamefont{Freiman}}, \bibinfo{journal}{Fizika Nizkikh Temperatur}
  \textbf{\bibinfo{volume}{36}}, \bibinfo{pages}{423} (\bibinfo{year}{2010}),
  \bibinfo{note}{[Low. Temp. Phys. 36, 333 (2010)]}.

\bibitem[{\citenamefont{Zha et~al.}(2004)\citenamefont{Zha, Mao, and
  Hemley}}]{zha04prb70:174107}
\bibinfo{author}{\bibfnamefont{C.-S.} \bibnamefont{Zha}},
  \bibinfo{author}{\bibfnamefont{H.-k.} \bibnamefont{Mao}}, \bibnamefont{and}
  \bibinfo{author}{\bibfnamefont{R.~J.} \bibnamefont{Hemley}},
  \bibinfo{journal}{Phys. Rev. B} \textbf{\bibinfo{volume}{70}},
  \bibinfo{pages}{174107} (\bibinfo{year}{2004}).

\bibitem[{\citenamefont{Takemura}(2001)}]{takemura01jap89:662}
\bibinfo{author}{\bibfnamefont{K.}~\bibnamefont{Takemura}},
  \bibinfo{journal}{J. Appl. Phys.} \textbf{\bibinfo{volume}{89}},
  \bibinfo{pages}{662} (\bibinfo{year}{2001}).

\bibitem[{\citenamefont{Zha et~al.}(1993)\citenamefont{Zha, Duffy, Mao, and
  Hemley}}]{zha93prb48:9246}
\bibinfo{author}{\bibfnamefont{C.-S.} \bibnamefont{Zha}},
  \bibinfo{author}{\bibfnamefont{T.~S.} \bibnamefont{Duffy}},
  \bibinfo{author}{\bibfnamefont{H.-k.} \bibnamefont{Mao}}, \bibnamefont{and}
  \bibinfo{author}{\bibfnamefont{R.~J.} \bibnamefont{Hemley}},
  \bibinfo{journal}{Phys. Rev. B} \textbf{\bibinfo{volume}{48}},
  \bibinfo{pages}{9246} (\bibinfo{year}{1993}).

\bibitem[{\citenamefont{Freiman et~al.}(2015)\citenamefont{Freiman, Grechnev,
  Tretyak, Goncharov, and Gregoryanz}}]{freiman15fnt41:571}
\bibinfo{author}{\bibfnamefont{Yu.~A.} \bibnamefont{Freiman}},
  \bibinfo{author}{\bibfnamefont{A.}~\bibnamefont{Grechnev}},
  \bibinfo{author}{\bibfnamefont{S.~M.} \bibnamefont{Tretyak}},
  \bibinfo{author}{\bibfnamefont{A.~F.} \bibnamefont{Goncharov}},
  \bibnamefont{and}
  \bibinfo{author}{\bibfnamefont{E.}~\bibnamefont{Gregoryanz}},
  \bibinfo{journal}{Fizika Nizkikh Temperatur} \textbf{\bibinfo{volume}{41}},
  \bibinfo{pages}{571} (\bibinfo{year}{2015}),
  \bibinfo{note}{[arXiv:1505.02949]}.

\bibitem[{\citenamefont{Nabi et~al.}(2005)\citenamefont{Nabi, Vitos, Johansson,
  and Ahuja}}]{nabi05prb72:172102}
\bibinfo{author}{\bibfnamefont{Z.}~\bibnamefont{Nabi}},
  \bibinfo{author}{\bibfnamefont{L.}~\bibnamefont{Vitos}},
  \bibinfo{author}{\bibfnamefont{B.}~\bibnamefont{Johansson}},
  \bibnamefont{and} \bibinfo{author}{\bibfnamefont{R.}~\bibnamefont{Ahuja}},
  \bibinfo{journal}{Phys. Rev. B} \textbf{\bibinfo{volume}{72}},
  \bibinfo{pages}{172102} (\bibinfo{year}{2005}).

\bibitem[{\citenamefont{Steinle-Neumann
  et~al.}(1999)\citenamefont{Steinle-Neumann, Stixrude, and
  Cohen}}]{steinle99prb60:791}
\bibinfo{author}{\bibfnamefont{G.}~\bibnamefont{Steinle-Neumann}},
  \bibinfo{author}{\bibfnamefont{L.}~\bibnamefont{Stixrude}}, \bibnamefont{and}
  \bibinfo{author}{\bibfnamefont{R.~E.} \bibnamefont{Cohen}},
  \bibinfo{journal}{Phys. Rev. B} \textbf{\bibinfo{volume}{60}},
  \bibinfo{pages}{791} (\bibinfo{year}{1999}).

\bibitem[{\citenamefont{McMahon et~al.}(2012)\citenamefont{McMahon, Morales,
  Pierleoni, and Ceperley}}]{mcmahon12rmp84:1607}
\bibinfo{author}{\bibfnamefont{J.~M.} \bibnamefont{McMahon}},
  \bibinfo{author}{\bibfnamefont{M.~A.} \bibnamefont{Morales}},
  \bibinfo{author}{\bibfnamefont{C.}~\bibnamefont{Pierleoni}},
  \bibnamefont{and} \bibinfo{author}{\bibfnamefont{D.~M.}
  \bibnamefont{Ceperley}}, \bibinfo{journal}{Rev. Mod. Phys.}
  \textbf{\bibinfo{volume}{84}}, \bibinfo{pages}{1607} (\bibinfo{year}{2012}).

\bibitem[{\citenamefont{Cazorla and Boronat}(2015)}]{cazorla15prb91:024103}
\bibinfo{author}{\bibfnamefont{C.}~\bibnamefont{Cazorla}} \bibnamefont{and}
  \bibinfo{author}{\bibfnamefont{J.}~\bibnamefont{Boronat}},
  \bibinfo{journal}{Phys. Rev. B} \textbf{\bibinfo{volume}{91}},
  \bibinfo{pages}{024103} (\bibinfo{year}{2015}).

\bibitem[{\citenamefont{Monserrat et~al.}(2014)\citenamefont{Monserrat,
  Drummond, Pickard, and Needs}}]{monserrat14prl112:055504}
\bibinfo{author}{\bibfnamefont{B.}~\bibnamefont{Monserrat}},
  \bibinfo{author}{\bibfnamefont{N.~D.} \bibnamefont{Drummond}},
  \bibinfo{author}{\bibfnamefont{C.~J.} \bibnamefont{Pickard}},
  \bibnamefont{and} \bibinfo{author}{\bibfnamefont{R.~J.} \bibnamefont{Needs}},
  \bibinfo{journal}{Phys. Rev. Lett.} \textbf{\bibinfo{volume}{112}},
  \bibinfo{pages}{055504} (\bibinfo{year}{2014}).

\bibitem[{\citenamefont{Wills et~al.}(2010)\citenamefont{Wills, Eriksson,
  Andersson, Delin, Grechnyev, and Alouani}}]{wills:rspt-book}
\bibinfo{author}{\bibfnamefont{J.~M.} \bibnamefont{Wills}},
  \bibinfo{author}{\bibfnamefont{O.}~\bibnamefont{Eriksson}},
  \bibinfo{author}{\bibfnamefont{P.}~\bibnamefont{Andersson}},
  \bibinfo{author}{\bibfnamefont{A.}~\bibnamefont{Delin}},
  \bibinfo{author}{\bibfnamefont{O.}~\bibnamefont{Grechnyev}},
  \bibnamefont{and} \bibinfo{author}{\bibfnamefont{M.}~\bibnamefont{Alouani}},
  \emph{\bibinfo{title}{Full-Potential Electronic Structure Method: Energy and
  Force Calculations with Density Functional and Dynamical Mean Field Theory}}
  (\bibinfo{publisher}{Springer}, \bibinfo{year}{2010}).

\bibitem[{\citenamefont{Perdew et~al.}(1996)\citenamefont{Perdew, Burke, and
  Ernzerhof}}]{perdew96prl77:3865}
\bibinfo{author}{\bibfnamefont{J.~P.} \bibnamefont{Perdew}},
  \bibinfo{author}{\bibfnamefont{K.}~\bibnamefont{Burke}}, \bibnamefont{and}
  \bibinfo{author}{\bibfnamefont{M.}~\bibnamefont{Ernzerhof}},
  \bibinfo{journal}{Phys. Rev. Lett.} \textbf{\bibinfo{volume}{77}},
  \bibinfo{pages}{3865} (\bibinfo{year}{1996}).

\bibitem[{\citenamefont{Freiman and Tretyak}(2007)}]{freiman07fnt33:719}
\bibinfo{author}{\bibfnamefont{Yu.~A.} \bibnamefont{Freiman}} \bibnamefont{and}
  \bibinfo{author}{\bibfnamefont{S.~M.} \bibnamefont{Tretyak}},
  \bibinfo{journal}{Fizika Nizkikh Temperatur} \textbf{\bibinfo{volume}{33}},
  \bibinfo{pages}{719} (\bibinfo{year}{2007}), \bibinfo{note}{[Low. Temp. Phys.
  33, 545 (2007)]}.

\bibitem[{\citenamefont{Freiman et~al.}(2008)\citenamefont{Freiman, Goncharov,
  Tretyak, Grechnev, Tse, Errandonea, Mao, and Hemley}}]{freiman08prb78:014301}
\bibinfo{author}{\bibfnamefont{Yu.~A.} \bibnamefont{Freiman}},
  \bibinfo{author}{\bibfnamefont{A.~F.} \bibnamefont{Goncharov}},
  \bibinfo{author}{\bibfnamefont{S.~M.} \bibnamefont{Tretyak}},
  \bibinfo{author}{\bibfnamefont{A.}~\bibnamefont{Grechnev}},
  \bibinfo{author}{\bibfnamefont{J.~S.} \bibnamefont{Tse}},
  \bibinfo{author}{\bibfnamefont{D.}~\bibnamefont{Errandonea}},
  \bibinfo{author}{\bibfnamefont{H.-k.} \bibnamefont{Mao}}, \bibnamefont{and}
  \bibinfo{author}{\bibfnamefont{R.~J.} \bibnamefont{Hemley}},
  \bibinfo{journal}{Phys. Rev. B} \textbf{\bibinfo{volume}{78}},
  \bibinfo{pages}{014301} (\bibinfo{year}{2008}).

\bibitem[{\citenamefont{Freiman et~al.}(2013)\citenamefont{Freiman, Grechnev,
  Tretyak, Goncharov, Zha, and Hemley}}]{freiman13prb88:214501}
\bibinfo{author}{\bibfnamefont{Yu.~A.} \bibnamefont{Freiman}},
  \bibinfo{author}{\bibfnamefont{A.}~\bibnamefont{Grechnev}},
  \bibinfo{author}{\bibfnamefont{S.~M.} \bibnamefont{Tretyak}},
  \bibinfo{author}{\bibfnamefont{A.~F.} \bibnamefont{Goncharov}},
  \bibinfo{author}{\bibfnamefont{C.~S.} \bibnamefont{Zha}}, \bibnamefont{and}
  \bibinfo{author}{\bibfnamefont{R.~J.} \bibnamefont{Hemley}},
  \bibinfo{journal}{Phys. Rev. B} \textbf{\bibinfo{volume}{88}},
  \bibinfo{pages}{214501} (\bibinfo{year}{2013}).

\bibitem[{\citenamefont{Barron and Klein}(1965)}]{barron65pps85:223}
\bibinfo{author}{\bibfnamefont{T.~H.~K.} \bibnamefont{Barron}}
  \bibnamefont{and} \bibinfo{author}{\bibfnamefont{M.~L.} \bibnamefont{Klein}},
  \bibinfo{journal}{Proc. Phys. Soc.} \textbf{\bibinfo{volume}{85}},
  \bibinfo{pages}{523} (\bibinfo{year}{1965}).

\bibitem[{\citenamefont{Watt and Peselnick}(1980)}]{watt80jap51:1525}
\bibinfo{author}{\bibfnamefont{J.~P.} \bibnamefont{Watt}} \bibnamefont{and}
  \bibinfo{author}{\bibfnamefont{L.}~\bibnamefont{Peselnick}},
  \bibinfo{journal}{J. Appl. Phys.} \textbf{\bibinfo{volume}{51}},
  \bibinfo{pages}{1525} (\bibinfo{year}{1980}).

\bibitem[{\citenamefont{Huang}(1950)}]{huang50prs203:178}
\bibinfo{author}{\bibfnamefont{K.}~\bibnamefont{Huang}},
  \bibinfo{journal}{Proc. Roy. Soc.} \textbf{\bibinfo{volume}{203}},
  \bibinfo{pages}{178} (\bibinfo{year}{1950}).

\bibitem[{\citenamefont{Mehl}(1993)}]{mehl93prb47:2493}
\bibinfo{author}{\bibfnamefont{M.~J.} \bibnamefont{Mehl}},
  \bibinfo{journal}{Phys. Rev. B} \textbf{\bibinfo{volume}{47}},
  \bibinfo{pages}{2493} (\bibinfo{year}{1993}).

\bibitem[{\citenamefont{Vinet et~al.}(1987)\citenamefont{Vinet, Smith,
  Ferrante, and Rose}}]{vinet87prb35:1945}
\bibinfo{author}{\bibfnamefont{P.}~\bibnamefont{Vinet}},
  \bibinfo{author}{\bibfnamefont{J.~R.} \bibnamefont{Smith}},
  \bibinfo{author}{\bibfnamefont{J.}~\bibnamefont{Ferrante}}, \bibnamefont{and}
  \bibinfo{author}{\bibfnamefont{J.~H.} \bibnamefont{Rose}},
  \bibinfo{journal}{Phys. Rev. B} \textbf{\bibinfo{volume}{35}},
  \bibinfo{pages}{1945} (\bibinfo{year}{1987}).

\bibitem[{\citenamefont{Olijnyk and Jephcoat}(2000)}]{olijnyk00jpcm12:10423}
\bibinfo{author}{\bibfnamefont{H.}~\bibnamefont{Olijnyk}} \bibnamefont{and}
  \bibinfo{author}{\bibfnamefont{A.~P.} \bibnamefont{Jephcoat}},
  \bibinfo{journal}{J. Phys.: Condens. Matter} \textbf{\bibinfo{volume}{12}},
  \bibinfo{pages}{10423} (\bibinfo{year}{2000}).

\bibitem[{\citenamefont{Musgrave}(1970)}]{musgrave:book}
\bibinfo{author}{\bibfnamefont{M.~J.~P.} \bibnamefont{Musgrave}},
  \emph{\bibinfo{title}{Crystal Acoustics}} (\bibinfo{publisher}{Holden-Day,
  San Francisco}, \bibinfo{year}{1970}).

\bibitem[{\citenamefont{Khairallah and
  Militzer}(2008)}]{khairallah08prl101:106407}
\bibinfo{author}{\bibfnamefont{S.~A.} \bibnamefont{Khairallah}}
  \bibnamefont{and} \bibinfo{author}{\bibfnamefont{B.}~\bibnamefont{Militzer}},
  \bibinfo{journal}{Phys. Rev. Lett.} \textbf{\bibinfo{volume}{101}},
  \bibinfo{pages}{106407} (\bibinfo{year}{2008}).

\bibitem[{\citenamefont{Nieto et~al.}(2012)\citenamefont{Nieto, Benedek, and
  Toennies}}]{nieto12njp14:013007}
\bibinfo{author}{\bibfnamefont{P.}~\bibnamefont{Nieto}},
  \bibinfo{author}{\bibfnamefont{G.}~\bibnamefont{Benedek}}, \bibnamefont{and}
  \bibinfo{author}{\bibfnamefont{J.~P.} \bibnamefont{Toennies}},
  \bibinfo{journal}{New J. Phys} \textbf{\bibinfo{volume}{14}},
  \bibinfo{pages}{013007} (\bibinfo{year}{2012}).

\end{thebibliography}

\end{document}